\newif\ifarxiv
\arxivtrue

\ifarxiv
\documentclass[acmsmall,nonacm]{acmart}
\else
\documentclass[acmsmall,screen]{acmart} 
\fi

\setcopyright{rightsretained}
\acmPrice{}
\acmDOI{10.1145/3591289}
\acmYear{2023}
\copyrightyear{2023}
\acmSubmissionID{pldi23main-p475-p}
\acmJournal{PACMPL}
\acmVolume{7}
\acmNumber{PLDI}
\acmArticle{175}
\acmMonth{6}
\ifarxiv \else
\received{2022-11-10}
\received[accepted]{2023-03-31}
\fi

\usepackage{booktabs}   
\usepackage{subcaption} 
\usepackage{prooftree}
\usepackage[shortlabels]{enumitem}
\usepackage{listings}
\usepackage{hhline}

\newcommand{\figref}[1]{\figurename~\ref{fig:#1}}
\newcommand{\secref}[1]{Sec.~\ref{sec:#1}}
\newcommand{\appref}[1]{App.~\ref{app:#1}}

\newcommand{\code}[1]{\ensuremath{\mathtt{#1}}}

\makeatletter

\newcommand{\multOpSym}{\ensuremath{\odot}}
\newcommand{\mult}{\ensuremath{\@ifnextchar\bgroup{\multT}{\multOpSym}}}
\newcommand{\multT}[2]{\ensuremath{\code{{#2}}}}

\newskip \point

\setbox136=\hbox{\leavevmode\raise0\point\hbox{${\lfloor}\kern-3\point{\lfloor}$}}
\setbox137=\hbox{\raise0\point\hbox{${ \rfloor}\kern-3\point{ \rfloor}$}}

\makeatother

\newcommand{\peter}[1] {{#1}}
\newcommand{\thibault}[1] {{#1}}

\makeatletter
\def\operator#1{\@ifnextchar\bgroup {\operatorarg{\ensuremath{#1}}}{\ensuremath{#1}}}
\def\operatorarg#1#2{{#1}{\ensuremath{(#2)}}}
\def\spoperator#1{\@ifnextchar\bgroup{\spoperatorarg{\ensuremath{#1}}}{\ensuremath{#1}}}
\def\spoperatorarg#1#2{\ensuremath{#1~#2}}

\def\method#1{\@ifnextchar\bgroup{\methodarg{\code{#1}}}{\code{#1}\xspace}}
\def\methodarg#1#2{{#1}\code{(#2)}\xspace}

\def\dom{\operator{\textit{dom}}}

\makeatother

\def\trueval{\textit{true}}

\newcommand\stackCloserHelp[2]{{%
    \setbox0\hbox{\ensuremath{#1}}%
    \rlap{\hbox to \wd0{\hss\raisebox{-.7\height}{#2}\hss}}\box0
}}

\makeatletter

\newcommand\msminus{\ensuremath{\setminus^{\#}}}

\newcommand{\marco}[1]{#1}
\newcommand{\marcon}[1]{#1}
\newcommand\freevars[1]{\ensuremath{\mathit{fv}(#1)}}

\newcommand\noop{\ensuremath{\mathtt{skip}}}

\newcommand\abort{\ensuremath{\mathtt{abort}}}

\newcommand\seq[2]{\ensuremath{#1;#2}}

\newcommand\assign[2]{\ensuremath{#1{:}{=}#2}}
\newcommand\cread[2]{\ensuremath{\assign{#1}{[#2]}}}
\newcommand\cwrite[2]{\ensuremath{\assign{[#1]}{#2}}}
\newcommand\alloc[2]{\ensuremath{\assign{#1}{\mathtt{alloc}(#2)}}}

\newcommand\cond[3]{\ensuremath{\mathtt{if~} (#1) \mathtt{~then~} \{#2\} \mathtt{~else~} \{#3\}}}

\newcommand\while[2]{\ensuremath{\mathtt{while~} (#1) \mathtt{~do~} \{#2\}}}
\newcommand\whileml[2]{\ensuremath{\noindent \displaystyle \begin{array}[t]{l}
    \mathtt{while~} (#1) \mathtt{~do~} \{ \\
    ~~~~#2\\
    \}\\
  \end{array}
}}

\newcommand\logicName{\textsc{CommCSL}\xspace}
\newcommand\toolName{\textsc{HyperViper}\xspace}
\newcommand\aguard[3]{\ensuremath{\mathtt{guard}_{#1}(#2, #3)}}

\newcommand\below[2]{\ensuremath{#2(#1)}}

\newcommand\pointsto[3]{\ensuremath{#1 \mapsto^{#2} #3}}

\newcommand\mappredname{\mathit{Map}}

\newcommand\mappred[2]{\ensuremath{\mappredname(#1, #2)}}
\newcommand\listpred[2]{\ensuremath{\mathit{List}(#1, #2)}}
\newcommand\fdomain[1]{\ensuremath{\mathit{dom}(#1)}}
\newcommand\llow{\ensuremath{\textsc{Low}}}

\newcommand\seqtoms[1]{\ensuremath{\mathit{ms}(#1)}}
\newcommand\msliteral[1]{\ensuremath{\{#1\}^{\#}}}
\newcommand\msadd{\ensuremath{\cup^{\#}}}
\newcommand\msunion[2]{\ensuremath{#1 \cup^{\#} #2}}
\newcommand\aguardlbl[4]{\ensuremath{\mathit{sguard}^{#3}(#2, #4)}}
\newcommand\aguardnolbl[3]{\ensuremath{\mathit{uguard}_{#2}(#3)}}

\newcommand\noguard[1]{\ensuremath{\mathit{noguard}(#1)}}
\newcommand\guheapsingle[2]{\ensuremath{[#2 \mapsto #1]}}

\newcommand\mmtriple[4]{\ensuremath{#1 \vdash \{ #2 \} #3 \{ #4 \}}}
\newcommand\mmttriple[4]{\ensuremath{#1 \models \{ #2 \} #3 \{ #4 \}}}
\newcommand\mmconf[3]{\ensuremath{\langle #1, (#2, #3) \rangle}}
\newcommand\mmstepsto[2]{\ensuremath{#1 \rightarrow #2}}
\newcommand\mmstepstox[2]{\ensuremath{#1 \rightarrow^* #2}}
\newcommand\mmeval[2]{\ensuremath{ \llbracket #1 \rrbracket (#2) }}

\newcommand\parc[2]{\ensuremath{#1 || #2}}

\newcommand\atomic[1]{\ensuremath{\mathtt{atomic}\mathtt{~}#1}}

\newcommand\emp{\ensuremath{\mathtt{emp}}}

\newcommand\store{\ensuremath{s}}

\newcommand\pstate{\ensuremath{\sigma}}
\newcommand\statefunc{\ensuremath{\Psi}}
\newcommand\normalize[1]{\ensuremath{\mathit{norm}(#1)}}

\newcommand\stepstox[4]{\ensuremath{\langle #1, #2 \rangle \rightarrow^{*} \langle #3, #4 \rangle}}

\newcommand\expval[2]{\ensuremath{[\!\![ #1 ]\!\!]_{#2}}}

\newcommand\sharedactions{\ensuremath{\mathit{SharedActions}}}
\newcommand\uniqueactions{\ensuremath{\mathit{UniqueActions}}}

\newcommand\nheap{\ensuremath{h}}
\newcommand\pheap{\ensuremath{\mathit{ph}}}
\newcommand\gheap{\ensuremath{\mathit{gh}}}
\newcommand\gheapf{\ensuremath{\mathit{\widetilde{gh}}}}
\newcommand\gheapfg{\ensuremath{\mathit{\widehat{gh}}}}
\newcommand\gsheap{\ensuremath{\mathit{gs}}}
\newcommand\guheap{\ensuremath{\mathit{gu}}}
\newcommand\guheaps{\ensuremath{\mathit{Gu}}}

\newcommand\modset[1]{\ensuremath{\mathit{mod}(#1)}}

\newcommand\rightv[1]{\ensuremath{\mathit{Right}(#1)}}

\newcommand\leftv[1]{\ensuremath{\mathit{Left}(#1)}}

\newcommand\unaryass[1]{\ensuremath{\mathit{unary}\text{~}#1}}
\newcommand\funcdomain[1]{\ensuremath{\mathit{dom}(#1)}}
\newcommand\funcdom[1]{\ensuremath{\mathit{dom}(#1)}}

\newcommand\locktype[2]{\ensuremath{\mathit{type}_{#1}(#2)}}

\newcommand\lockabstr[2]{\ensuremath{\alpha_{#1}(#2)}}

\newcommand\lockpre[2]{\ensuremath{\mathit{pre}_{#1}(#2)}}
\newcommand\allpre[2]{\ensuremath{\mathit{PRE}_{#1}(#2)}}

\makeatletter
\newcommand{\oset}[3][0ex]{%
  {\mathop{#3}\limits^{
    \vbox to#1{\kern-2\ex@
    \hbox{$\scriptscriptstyle#2$}\vss}}}}
\makeatother

\newcommand\new[2]{\@ifnextchar\bgroup{\newthree{#1}{#2}}{\newthree{#1}{#2}{}}}

\newcommand{\triple}[3]{\@ifnextchar\bgroup{\tripleFour{#1}{#2}{#3}}{\tripleFour{\env}{#1}{#2}{#3}}}
\newcommand{\tripleFour}[4]{\ensuremath{{#1}\vdash \{#2\}~#3~\{#4\}}}
\newcommand{\env}{\Lambda}

\newcommand{\emptymap}{\ensuremath{\mathit{empty}}}
\newcommand{\emptymultiset}{\ensuremath{\emptyset^{\#}}}

\usepackage{relsize}

\makeatletter

\newcommand{\judgementdef}[3]{\ensuremath{#2#1 #3}}
\newcommand{\judgement}[1]{\@ifnextchar\bgroup{\judgementArg{#1}}{\judgementdef{#1}{}{}}}
\newcommand{\judgementArg}[2]{\@ifnextchar\bgroup{\judgementArgs{#1}{#2}}{\judgementdef{#1}{}{#2}}}
\newcommand{\judgementArgs}[3]{\@ifnextchar\bgroup{#2,\judgementArgs{#1}{#3}}{\judgementdef{#1}{#2}{#3}}}

\newcommand{\OKsym}{\ensuremath{\vdash_{\textit{OK}}}}
\newcommand{\OK}{\@ifnextchar\bgroup{\OKone}{\ensuremath{\OKsym}}}
\newcommand{\OKone}[1]{\@ifnextchar\bgroup{\OKmore{#1}}{\ensuremath{#1 \OKsym}}}
\newcommand{\OKmore}[2]{\@ifnextchar\bgroup{\OKthree{#1,#2}}{\OKone{#1,#2}}}
\newcommand{\OKthree}[3]{\ensuremath{#1,#2\OKsym #3}}

\newcommand{\futureEntailsSym}{\textit{futureEntails}}
\newcommand{\futureEntails}{\@ifnextchar\bgroup{\futureEntailsArg}{\futureEntailsSym}}
\newcommand{\futureEntailsArg}[3]{\@ifnextchar\bgroup{\futureEntailsArg{#1,#2}{#3}}{\ensuremath{(#1){.}\futureEntailsSym(#2,#3)}}}

\newcommand{\futureCombinesSym}{\textit{futureCombines}}
\newcommand{\futureCombines}{\@ifnextchar\bgroup{\futureCombinesArg}{\futureCombinesSym}}
\newcommand{\futureCombinesArg}[4]{\@ifnextchar\bgroup{\futureCombinesArg{#1,#2}{#3}{#4}}{\ensuremath{(#1){.}\futureCombinesSym(#2,#3,#4)}}}

\newcommand{\freeze}{\@ifnextchar\bgroup{\freezeone}{\ensuremath{\textit{freeze}}}}
\newcommand{\freezeone}[1]{\@ifnextchar\bgroup{\freezemore{#1}}{\ensuremath{\textit{freeze}(#1)}}}
\newcommand{\freezemore}[2]{\freeze{#1,#2}}

\def\call#1#2{\@ifnextchar\bgroup{\calldef{#1}{#2}}{\calldef{#1}{#2}{}}} 
\def\calldef#1#2#3{\@ifnextchar\bgroup{\callmergeargs{#1}{#2}{#3}}{\deref{#1}{{#2}({#3})}}}
\def\callmergeargs#1#2#3#4{\calldef{#1}{#2}{#3,#4}}
\def\deref#1#2{\ensuremath{#1{.}#2}}

\newcommand{\response}[2]{\@ifnextchar\bgroup{\responsethree{#1}{#2}}{\call{#1}{#2}}}
\newcommand{\responsethree}[3]{\@ifnextchar\bgroup{\responsewhere{#1}{#2}{#3}}{\responsewhere{#1}{#2}{}{#3}}} 
\newcommand{\responsewhere}[4]{\call{#1}{#2}{#3}\;\where\;{#4}}
\newcommand{\where}{\textit{where}}

\usepackage{xspace}

\ifx\symlasy\undefined   \DeclareSymbolFont{lasy}{U}{lasy}{m}{n}
  \SetSymbolFont{lasy}{bold}{U}{lasy}{b}{n} \else
\fi
\DeclareMathSymbol\safeleadsto {\mathrel}{lasy}{"3B}

\newskip \point

\setbox136=\hbox{\leavevmode\raise0\point\hbox{${\lfloor}\kern-3\point{\lfloor}$}}
\setbox137=\hbox{\raise0\point\hbox{${ \rfloor}\kern-3\point{ \rfloor}$}}

\def \premisespacing{\quad\;}

\newcommand{\rulename}[1]{\ensuremath{(\textit{#1})}}

\def \RulePremisesNewlineMore[#1]#2.#3#4{\@ifnextchar\bgroup{\RulePremisesNewlineMore[#1]{#2}.{#3\premisespacing#4}}{\@ifnextchar.{\RulePremisesNewline[#1]{{\begin{array}{c}#2\\#3\premisespacing#4\end{array}}}}{\RuleMultiPremise[#1]{{\begin{array}{c}#2\\#3\end{array}}}{#4}}}}

\def \RulePremisesNewline[#1]#2.#3{\@ifnextchar\bgroup{\RulePremisesNewlineMore[#1]{#2}.{#3}}{\@ifnextchar.{\RulePremisesNewline[#1]{{\begin{array}{c}#2\\#3\end{array}}}}{\RuleMultiPremise[#1]{#2}{#3}}}}

\def \RuleMultiPremise[#1]#2#3{\@ifnextchar\bgroup{\RuleMultiPremise[#1]{#2\premisespacing#3}}{\@ifnextchar.{\RulePremisesNewline[#1]{#2\premisespacing#3}}{\prooftree #2\justifies#3 \using{#1}\endprooftree}}}

\def \RuleWithName[#1]#2{\@ifnextchar\bgroup {\RuleMultiPremise[#1]{#2}}{\@ifnextchar.{\RulePremisesNewline[#1]{#2}}{\prooftree \justifies #2 \using{#1} \endprooftree}}}

\def \RuleWithInfo[#1]{\@ifnextchar[{\RuleWithNameAndCondition[#1]}{\RuleWithName[(\textsc{#1})]}}

\def \RuleWithNameAndCondition[#1][#2]{\RuleWithName[\rulename{#1}^{#2}]}

\def \Inf{\proofrulebaseline=2ex \abovedisplayskip12\point\belowdisplayskip12\point \abovedisplayshortskip8\point\belowdisplayshortskip8\point \@ifnextchar[{\RuleWithInfo}{\RuleWithName[ ]}}

\makeatother

\usepackage{pstricks}
\def \longharpoon#1{\psset{unit=1\point,linewidth=0.35\point}%
\psline{cc-cc}(0,1.5)(.9#1,1.5)%
\hspace*{.9#1}\pscurve{cc-cc}(0,1.5)(-1.1,2)(-2.25,3.7)\hspace*{.1#1}}

\def \Vec#1{\setbox155=\hbox{$#1$}%
\leavevmode\copy155\kern-.95\wd155
\raise\ht155\hbox{\longharpoon{\wd155}}}

\newcommand{\appendixref}[1]{\ifarxiv \appref{#1} \else \citet{commcslarxiv} \fi}

\begin{document}

\title{\logicName: Proving Information Flow Security for Concurrent Programs using Abstract Commutativity}         

\author{Marco Eilers}
\orcid{0000-0003-4891-6950}             
\affiliation{
  \department{Department of Computer Science}              
  \institution{ETH Zurich}            
  \city{Zurich}
  \country{Switzerland}                    
}
\email{marco.eilers@inf.ethz.ch}          

\author{Thibault Dardinier}
\orcid{0000-0003-2719-4856}             
\affiliation{
  \department{Department of Computer Science}              
  \institution{ETH Zurich}            
  \city{Zurich}
  \country{Switzerland}                  
}
\email{thibault.dardinier@inf.ethz.ch}         

\author{Peter M\"uller}
\orcid{0000-0001-7001-2566}             
\affiliation{
  \department{Department of Computer Science}              
  \institution{ETH Zurich}            
  \city{Zurich}
  \country{Switzerland}                 
}
\email{peter.mueller@inf.ethz.ch}         

\begin{abstract}
Information flow security ensures that the secret data manipulated by a program does not influence its observable output. Proving information flow security is especially challenging for concurrent programs, where operations on secret data may influence the execution time of a thread and, thereby, the interleaving between threads. Such \emph{internal timing channels} may affect the observable outcome of a program even if an attacker does not observe execution times. Existing verification techniques for information flow security in concurrent programs attempt to prove that secret data does not influence the relative timing of threads. However, these techniques are often restrictive (for instance because they disallow branching on secret data) and make strong assumptions about the execution platform (ignoring caching, processor instructions with data-dependent execution time, and other common features that affect execution time).

In this paper, we present a novel verification technique for secure information flow in concurrent programs that lifts these restrictions and does not make any assumptions about timing behavior. The key idea is to prove that all mutating operations performed on shared data commute, such that different thread interleavings do not influence its final value. Crucially, commutativity is required only for an \emph{abstraction} of the shared data that contains the information that will be leaked to a public output. Abstract commutativity is satisfied by many more operations than standard commutativity, which makes our technique widely applicable.

We formalize our technique in \logicName, a relational concurrent separation logic with support for commutativity-based reasoning, and prove its soundness in Isabelle/HOL\@. We have implemented \logicName in \toolName, an automated verifier based on the Viper verification infrastructure, and demonstrate its ability to verify challenging examples.
\end{abstract}



\maketitle

\section{Introduction}

Reasoning about information flow is important to ensure the confidentiality and integrity of data. \peter{The main goal of information flow security is to ensure the absence of \emph{value channels}, that is, unwanted information flows through the \emph{result values} computed by a program.} For sequential programs, preventing value channels is conceptually simple, and is enabled by various existing type systems, program logics, and static analyses.

\peter{Depending on the application scenario, it is sometimes desirable to also prevent side channels that let an attacker obtain information about secret values indirectly, by observing parameters such as execution time.} Preventing timing channels requires ruling out \emph{all possible} sources of secret-dependent timing differences, which may depend on complex \marco{(but standard)} hardware features like caching~\cite{DBLP:conf/esorics/StefanBYLTRM13} and speculative execution~\cite{spectre,meltdown}. 

For sequential programs, proving the absence of value channels does not involve reasoning about timing. Thus, for the many scenarios where attackers cannot observe execution time at all (e.g., because the time when data is made public is decoupled from its computation time, \peter{which is common for batch processing}) or not with sufficient precision (e.g., because execution time is hidden by a laggy network), \marco{simple} reasoning about values is sufficient \marco{even on standard hardware.}

However, \peter{this is not the case for} concurrent programs with shared memory, \peter{where} timing differences  \peter{may result in value channels, which can be observed even by attackers that cannot observe execution time. The  program in \figref{initexample} illustrates this problem}\thibault{.}

\begin{figure}
\centering
\begin{tabular}{cl}
\multicolumn{2}{c}{$\assign{t1, t2~}{~0}$}\\
\multicolumn{1}{l||}{$\whileml{t1 < 100}{\hspace{0.28cm}\assign{t1~}{~t1 + 1}}$} & $\whileml{t2 < h}{\hspace{0.28cm}\assign{t2~}{~t2 + 1}}$ \\
\multicolumn{1}{l||}{\hspace{0.18cm}$\assign{s}{3}$}                          & \hspace{0.18cm}$\assign{s}{4}$                        \\
\multicolumn{2}{c}{$\mathit{print}(s)$}                                                                  
\end{tabular}
\caption{Example program: Timing channels become value channels.} \label{fig:initexample}
\end{figure}

Since no information about the secret value $h$ is leaked to the output $s$ through direct assignments or explicit control flow, each thread individually does not leak secret information. 
However, the execution time of the loop in the right thread depends on the value of $h$. This timing may affect \thibault{the order in which} the threads perform their assignment to $s$ and, thus, which value is printed in the end. With a deterministic round-robin scheduler, the printed value leaks whether or not $h$ is greater than 100. With a non-deterministic scheduler with a known probability distribution, the entire value of $h$ is leaked probabilistically over multiple executions. 

\peter{Information leaks that turn secret-dependent timing differences between threads into observable differences in the program state have long been recognized and are called \emph{internal timing channels}~\cite{DBLP:conf/csfw/VolpanoS98}.
Some existing solutions to this problem essentially eliminate shared memory between threads~\cite{martin,joana}, which rules out many useful concurrent programming patterns. Others employ the techniques used to prevent standard timing channels. They forbid programs from including operations whose execution time depends on secret inputs~\cite{covern,veronica,prob,smith}; in our example, they would reject the loop in the right thread. However, these techniques typically assume idealized hardware, where secret-depending branching is the only source of timing channels. In principle, they can be extended to standard hardware (e.g., one commonly prevents timing leaks through cache effects by forbidding all secret-dependent memory accesses~\cite{itsnotthatdifficult}). 
However, attempting to rule out \emph{all} sources of timing differences requires information about compilers~\cite{DBLP:conf/csfw/BartheGL18} and the hardware on which code is to be executed~\cite{DBLP:journals/taco/CleemputCS12,DBLP:conf/ccs/AndryscoNBJS18}}, \marco{including hardware details manufacturers usually do not make public. As a result, it is virtually impossible to rule out timing channels with absolute certainty, and thus,
existing techniques \emph{cannot rule out value channels in concurrent programs on standard hardware with certainty} either, since they require eliminating timing channels.}

\paragraph{This work}
\peter{We present a novel technique to prove the absence of value channels} in concurrent programs that does \emph{not} require reasoning about timing \peter{and, thus, is sound even in the presence of complex modern hardware features. In the common case that attackers cannot observe execution time, our technique enables conceptually simple information flow security proofs.} When the attacker \emph{can} observe timing \marco{to a degree}, our technique can be combined with orthogonal techniques to \marco{try to} rule out timing channels.
Crucially, our technique is \emph{modular}, i.e., it does not require explicit reasoning about thread interleavings, which would make verification computationally infeasible.

\peter{Our technique is based on the observation that different, secret-dependent interleavings \marco{of shared data mutations} lead to the same final result}
if the mutating operations on shared data performed by different threads \emph{commute}. 
For instance, the program above is (rightfully) rejected by our technique because the two assignments to the shared variable $s$ do not commute; however, a variation of this example where the concurrent assignments are modified so that the left thread atomically assigns $s+3$ to $s$, and the right thread assigns $s+4$, is allowed: This program always \peter{increases $s$ by $7$}, since the additions performed by the two threads commute and their \peter{(secret-dependent)} order therefore does not influence the final result. \peter{Consequently, the adapted program contains no value channels.}

Standard commutativity is a strong requirement that is satisfied only by a minority of operations of typical data structures. A key insight of our work is to restrict the commutativity requirement to those parts of the shared data that are required to be non-secret (or \emph{public}). In our example, if $s$ were secret (and consequently not printed at the end of the program) then the program would be information flow secure even though the assignments to $s$ do not commute. To this end, we allow programmers to specify abstract views of shared data structures that contain the information that must remain public; all information not included in the abstraction is considered secret. It is then sufficient to prove \emph{abstract commutativity}, that is, the order of two operations does not affect the abstract view of a data structure, but it may affect other aspects. For instance, two list-append operations commute under a set- or length-abstraction of the list, but do not commute on the concrete list (unless they append the same element).

\peter{Many programs expose public views on data structures that contain secret information. One common case is maintaining} \emph{internal data} that intentionally contains secret information (e.g., an employee database with individual salaries), and then exposing only a part of it (e.g., average salaries for reporting purposes); in this case, our technique allows specifying the projection to averages as the abstraction.
\peter{Another common case is maintaining data that is \emph{intended} to be public, but inadvertently tainted} with secret information as a result of timing differences in a concurrent computation. \peter{For instance, a program may maintain a list of anonymized accounts, but the order in the list depends on the time it took to process the (secret) purchases of each account.}
In this case, we can use the multiset view of the list as the abstraction, and prove, for instance, that \peter{a sorted version of the list can be exposed without creating a value channel.}

We formalize our technique as a relational concurrent separation logic with special constructs for reasoning about abstract commutativity of concurrent operations. This logic, \logicName, enables \emph{modular} proofs of information flow security, does not require reasoning about the timing behavior of threads, and is amenable to automation via SMT solvers. Our logic supports an expressive assertion language that can express concepts like value-dependent secrecy~\cite{covern}. 
In addition to proving existing applications secure, the ideas behind our technique can also be used as a guiding principle when building concurrent applications that deal with secret data.

\paragraph{Contributions and outline}

We make the following contributions:

\begin{itemize}

\item We show how commutativity can be used to prove information flow security in shared-memory concurrent programs. Our verification technique allows threads to flexibly manipulate secret data and does not require reasoning (nor make assumptions) about the timing of executions, such that it is directly applicable to programs running on standard hardware.

\item We introduce the generalized notion of abstract commutativity, which requires operations to commute only relative to a user-defined \marcon{and application-specific} abstraction of the shared state. Abstract commutativity applies to a much wider set of common scenarios.

\item We incorporate these techniques into \logicName, a novel concurrent separation logic that enables modular proofs of information flow security using abstract commutativity.

\item We formalize our logic and prove its soundness in Isabelle/HOL~\cite{isabelle}.

\item We implement \logicName in an automated, SMT-based verification tool called \toolName based on the Viper verification infrastructure~\cite{viper}.

\item We evaluate our tool by proving several challenging examples information flow secure.
\end{itemize}

This paper is organized as follows: \secref{overview} provides an informal overview of our technique. We formally define \logicName in \secref{logic}, and summarize its soundness proof in \secref{soundness}. In \secref{impl}, we describe the implementation of the logic in \toolName and use it to verify several challenging programming patterns. We discuss related work in \secref{rw} and conclude in \secref{conclusion}.

\section{Overview}\label{sec:overview}

In this section, we explain the central concepts behind \logicName informally. \peter{We occasionally omit details for simplicity}; the full logic with all features and checks will be presented in \secref{logic}.

\subsection{Problem Statement}

Our goal in this paper is to prove that concurrent programs that have both high-sensitivity (high) and low-sensitivity (low) inputs do not leak information about the high inputs in their low output values\footnote{We limit our presentation to two security labels, high and low, instead of the general case of having arbitrary lattices of labels; however, techniques for verifying information flow security with two levels can be used to verify programs with arbitrary finite lattices by performing the verification multiple times, once for every element of the lattice.}. Formally, this can be expressed as \marcon{(termination-insensitive)} \emph{non-interference}~\cite{nonint}, a 2-safety \emph{hyperproperty}~\cite{hyperprop}, i.e., a property of pairs of finite execution traces of the program ($\store(x)$ denotes the value of variable $x$ in program store $\store$): 
 
\begin{definition}\label{def:ni}
  A program $c$ with a set of input variables $I$ and output variables $O$, of which some subsets $I_l \subseteq I$ and $O_l \subseteq O$ are low, 
  satisfies non-interference iff for all $\store_1, \store_2$ and $\store'_1, \store'_2$, 
  if $\forall x \in I_l \ldotp \store_1(x) = \store_2(x)$ and $\stepstox{c}{\store_1}{\noop}{\store'_1}$ 
  and $\stepstox{c}{\store_2}{\noop}{\store'_2}$, then $\forall x \in O_l \ldotp \store'_1(x) = \store'_2(x)$.
\end{definition}

\peter{This definition captures value channels.} As explained in the introduction, timing side channels \marco{can potentially be prevented} using orthogonal techniques if necessary.

\begin{figure}[t!]
\begin{minipage}[t]{0.43\textwidth}
\begin{lstlisting}
procedure targetSize(households) {
  // ensures: c is low
  n := |households|
  c := createCounter(0)
  (worker(households, 0, n/2, c) 
    || 
  worker(households, n/2, n, c))
  return c
}
\end{lstlisting}
\end{minipage}
\hfill
\begin{minipage}[t]{0.53\textwidth}
\begin{lstlisting}
procedure worker(households, f, t, c) {
  for (i in f..t-1) {
    targets := countTargets(household[i])
    atomic:
      c.add(targets)
  }
}

procedure countTargets(household: map[str, int]) {
  return household["nAdults"]
}
\end{lstlisting}
\end{minipage}
\vspace{-4mm}
\caption{Example program: Multiple threads add values to a shared counter. \code{households} is an array containing customer data per household. The result of  \code{countTargets} is low, but its execution time may be high, here due to potential hash collisions with other keys whose presence in the \code{households} map is secret.} \label{fig:setexample}
\end{figure}

The example in Fig.~\ref{fig:setexample} illustrates this property. The \code{targetSize} procedure determines the size of the target audience for a marketing campaign as the number of people in a given array of households that satisfy certain criteria. The procedure uses two worker threads that each iterate over half of the households, determine the number of target persons in each household, and add that number to a shared counter \code{c}. 
We assume that the number of households in the input as well as how many members of a household are in the target audience (i.e., the result value of \code{countTargets}) is low. 
However, the execution time of \code{countTargets} may depend on high data for a variety of reasons. \peter{For example, the implementation in Fig.~\ref{fig:setexample} consists of a simple look-up in a hash map. Nevertheless, its execution time may depend on hash collisions with other keys whose presence in the map is secret.}
Our goal is to prove that the output of \code{targetSize} is low.

\subsection{Commutativity-Based Information Flow Reasoning}

\peter{The high data accessed in \code{countTargets} may affect the execution time of the procedure and, thereby, the thread schedule and the intermediate values of the shared counter \code{c}. Due to this internal timing channel, the value of \code{c} \emph{must} be considered high \emph{during} the execution of the worker threads.}
However, \emph{after} the worker threads have terminated, the counter value can safely be considered to be low for two reasons. First, as per our assumption, each individual value added is low. Second, although high information may have affected the \emph{order} in which different values are added to the counter, this order does not affect the final counter value because the updates \emph{commute}.
This example illustrates the central insight of this paper: \emph{Internal timing channels do not affect the final outputs of a program if the modifications performed by different threads commute.}

Building on this insight, we present a verification technique for proving non-interference of concurrent programs. This technique enforces four central properties (in addition to standard checks that are sufficient to ensure non-interference for sequential programs):

\begin{enumerate}
  \item Low initial value: For any shared data structure that is to be modified by multiple threads, the initial contents of the data structure is low.
  \item Number of modifications is low: High data does not influence \marcon{if or how often} a thread performs a modification of the shared data at any given point in the program. 
  \item Modification arguments are low: Any data inserted into the shared data structure is low.
  \item Commutativity: The modifications performed by different threads commute with each other.
\end{enumerate}

These four properties are sufficient to ensure that, for any pair of executions of the verified program with identical low but potentially different high inputs, the \emph{final} value of any shared data structure is \marco{identical} after all threads have finished modifying it.

To understand why, consider two such executions: By property (1), the data structure will have the same contents in both executions when it is initially shared. By property (2), in both executions, the \emph{same number of modifications} will be performed on the data structure (e.g., in our example, the same number of values will be added). By property (3), for each modification in the first execution, there is a matching modification in the second execution that uses the same arguments (e.g., if the value $x$ has been added to the counter \marcon{$n$ times} in the first execution, then $x$ is also added \marcon{$n$ times} in the second execution); the only difference is in the order of modifications. However, by property (4), reordering the modifications leaves the final value unchanged. Consequently, the final contents of the data structure are the result of performing the same \peter{commutative} operations with the same arguments on the same initial value, and must therefore be identical.

\marcon{The basic idea behind our verification technique is thus to prove these four properties for every shared data structure, which then allows us to treat the \emph{final} contents of the data structure after concurrent modifications have finished to be low (whereas all \emph{intermediate} values read during concurrent modification will always have to be treated as high).}
In the rest of this section, we will build on and expand on this basic idea: We will relax some of the properties to make our verification technique more complete and more widely applicable, and we will explain how we check each of the four properties on the program to be verified.

\subsection{Abstract Commutativity}
\label{sec:abstract-commutativity}

\begin{figure}[t!]
\begin{minipage}[t]{0.53\textwidth}
\begin{lstlisting}
procedure targets(households) returns (res) {
  // ensures: res is low
  n := |households|
  m := createMap()
  (worker(households, 0, n/2, m) 
    || 
  worker(households, n/2, n, m))
  res := sort(toList(keys(m)))
}
\end{lstlisting}
\end{minipage}
\hfill
\begin{minipage}[t]{0.43\textwidth}
\begin{lstlisting}
procedure worker(households, f, t, m) {
  for (i in f..t-1) {
    adr, rsn := select(household[i])
    atomic:
      m.put(adr, rsn)
  }
}
\end{lstlisting}
\end{minipage}
\vspace{-4mm}
\caption{Example program: Multiple threads add values to a shared map.} \label{fig:mapkeyexample}
\end{figure}

\noindent
The methodology described so far is sound, but not sufficiently complete: 
First, it requires that \emph{all} information stored in shared data structures must be low after all concurrent modifications, even information that is never leaked to a public output. Second, most mutating operations on common data structures do not commute and therefore do not satisfy our property~(4) above.

As an example, consider a variation of the previous example in Fig.~\ref{fig:mapkeyexample}: Now, the shared data structure is a map, and each worker extracts a key-value pair per household, where the address (the key) is low, but the reason why an address was selected (the value)  is high.  Here, different invocations of \code{put} do not always commute: If two threads put the same key but different values, then the later put-operation will ``win'', and its value will overwrite the previous value for said key. As a result, the final contents of the map allow an observer to conclude which \code{put} happened later and, thereby, draw conclusions about the secrets that cause the different thread interleavings. 

However, the procedure \code{targets} does not actually return the entire contents of the map, but only (a sorted list representation of) its key set. Therefore, differences in values do not affect \code{targets}' public output. 
\peter{Scenarios where programs expose \marco{some} public views on the data they maintain are common in practice;}
to accommodate them, we allow programmers to define an \emph{abstract view} of the shared data structure that is guaranteed to be low and, therefore, allowed to be leaked. By focusing on the relevant part of a shared data structure, we no longer have to demand commutativity of all concurrent modifications, but only \emph{abstract commutativity}: commutativity modulo the abstract view. That is, we require that, if the \emph{abstract} view of the shared data is low initially, then switching the order of any two modifications on the data does not affect the final \emph{abstract} view of the data structure. This is fulfilled for the example in Fig.~\ref{fig:mapkeyexample}: Different \code{put}-operations do not commute w.r.t.\ the entire contents of the map, but they do commute w.r.t.\ the map's key set.
\marcon{Note that our abstraction does \emph{not} simply abstract away implementation details of the data structure to get a logical view of the data, but instead intentionally abstracts away all parts of the data structure that may contain high data, including otherwise vital parts of the data (like here the values in a map).}

\peter{Since we require only the abstract view of shared data to be low, we} no longer have to demand that \emph{all} arguments of map modifications have to be low: arguments that do not affect the abstract view of the data may contain high information. In the example, this allows us to insert high \emph{values} into the map as long as all inserted \emph{keys} are low. We use preconditions on mutating operations to specify which arguments must be low to ensure that the abstract view of the data remains low.

To summarize, we relax our four central properties as follows:
\begin{enumerate}
  \item Low initial \emph{abstract} value: The abstract view of the shared data is low when first shared.
  \item \marcon{Number of modifications is low}: Unchanged\footnote{One could in principle restrict property~(2) to modifications that affect the abstract view of the shared data structure. We omit this optimization since operations that do not affect the abstract value commute trivially and, therefore, do not complicate verification significantly.}.
  \item Modification arguments fulfill sufficient precondition: \peter{(a)}~Arguments of each modification fulfill a \peter{given} precondition that \peter{(b)}~ensures that the abstract view of the shared data remains low after the modification.
  \item \emph{Abstract} Commutativity: All modifications commute w.r.t.\ the abstract view of shared data. 
\end{enumerate}

By enabling these relaxations, abstract commutativity allows us to handle many more practical examples than standard commutativity,
\marcon{at the cost of weakening the resulting guarantee: Now we may consider only the abstraction of the final value of the data structure to be low after concurrent modifications have been performed (all its intermediate values read during concurrent modification must still be treated as high).}
To demonstrate that the relaxed conditions are sufficient,
we can make a similar argument as before to argue that the abstract view of each shared data structure will be low after all concurrent modifications on it have been performed.

\subsection{Resource Specifications}
\label{sec:resource-specification}

Proving our four properties \peter{directly on the level of} heap-manipulating programs is \peter{difficult}: For example, showing that two modifications commute (even without an abstraction) on actual program states means proving the equivalence of two programs that perform the modifications in a different order. If these operations  contain complex steps like pointer arithmetic, loops, or memory allocation (all of which might potentially be performed in real implementations of \code{put} methods on maps), it is non-trivial to even define \peter{equivalence of states} (e.g., because of non-deterministic allocation), and even more difficult to perform such proofs~\cite{DBLP:conf/vmcai/KoskinenB21}.

We therefore do not check all of our properties \emph{directly} on program states, but instead on pure mathematical values. We exploit the fact that, when working with separation logics, it is standard practice in the specification of data structures to use separation logic \emph{predicates} that relate the contents of data structures to pure values~\cite{ParkinsonBierman05}. For example, \peter{linked lists are typically specified using} a predicate of the form $\mathit{list}(p, s)$, where $p$ is the pointer to the start of the list and $s$ is a mathematical sequence describing the contents of the list; $\mathit{list}(p, s)$ holds iff $p$ points to a list whose contents are $s$. Methods that manipulate the data structures are then specified in terms of those pure values: for example, an \code{append} method of a  list implementation would usually require the list predicate in its precondition for some abstract value $s$, and return it in its postcondition, with an abstract value that \peter{was extended by} the appended value. 
We will assume that such specifications exist, and require additionally that the predicate \peter{uniquely determines the abstract value}, which is typically the case for existing definitions.  
Thus, we exploit existing standard verification constructs to map our used data structures to pure values, enabling us to check properties (3) and (4) on the level of those pure values.

To check property~(4), we must first determine which  modifications are performed on the shared data. Scanning the entire program for such modifications would not be modular. Instead, we associate a set of legal operations with a data structure when we share it, and subsequently enforce that all modifications that are performed by any thread correspond to one of the legal operations. 

\peter{We express these aspects of a shared data structure using a novel specification construct: a \emph{resource specification}} declares a \emph{pure} data type (\peter{for} the contents of a data structure in the standard separation logic style), an \emph{abstract view} in the form of an abstraction function $\alpha$ that maps a value of the pure data type to another \peter{mathematical} value that characterizes which aspects of the data structure ultimately have to be public, and a set of \emph{actions} that may be performed on the shared data structure. Each action comes with a function that defines how the action modifies the (pure) value of the data structure and with a \peter{(relational)} precondition that restricts the arguments of the action such that, after performing the action, the abstract view of the data structure remains low. 

Fig.~\ref{fig:examplelockspecmerged} (left) shows a resource specification for our map example: Its type is a partial mapping from keys to values, its only legal action is \textsc{Put}, which updates the mapping, and the precondition of \textsc{Put} requires the argument key (but not the value) to be low using an assertion which we will formally introduce later. We will discuss later what it means for an action to be shared or unique. 

\begin{figure}
\small
\begin{minipage}[t]{0.49\textwidth}
  	\begin{align*}
\locktype{\textsc{MK}}{v} &\equiv K \rightharpoonup V \\
\lockabstr{\textsc{MK}}{v} &\equiv \funcdomain{v} \\
	\sharedactions_{\textsc{MK}} &\equiv \{ \textsc{Put} \} \\
	\uniqueactions_{\textsc{MK}} &\equiv \emptyset \\
 	f_{\textsc{Put}}(v, \langle key, val \rangle) &\equiv m[k \mapsto v] \\
 	\lockpre{\textsc{Put}}{\langle key, val \rangle} &\equiv \below{key}{\llow}
	\end{align*}
\end{minipage}
\begin{minipage}[t]{0.49\textwidth}
	\begin{align*}
\locktype{\textsc{MD}}{v} &\equiv K \rightharpoonup V \\
\lockabstr{\textsc{MD}}{v} &\equiv v \\
	\sharedactions_{\textsc{MD}} &\equiv \emptyset \\
	\uniqueactions_{\textsc{MD}} &\equiv \{ \textsc{Put}_1, \textsc{Put}_2 \} \\
 	f_{\textsc{Put}_i}(v, \langle key, val \rangle) &\equiv m[k \mapsto v] \\
 	\lockpre{\textsc{Put}_i}{\langle key, val \rangle} &\equiv \below{key}{\llow} \wedge \below{val}{\llow} \\
 	&~~~  \wedge key \in \mathit{range}_i
	\end{align*}
\end{minipage}
\vspace{-4mm}
	\caption{Left: Complete resource specification for the map example in Fig.~\ref{fig:mapkeyexample}, where $K$ and $V$ are the types of the keys and values in the map, respectively. Right: Alternative resource specification that allows two different threads to  perform updates only in their own range of keys, which does not overlap with that of the other thread. We assume here that $\mathit{range}_1$ and $\mathit{range}_2$ do not overlap.}\label{fig:examplelockspecmerged}
\end{figure}

Crucially, such a resource specification is independent of any specific client program and of any specific implementation of a map data structure; it can be used in any program that uses a shared map whose key set is low in the end, and it can be combined with any separation logic predicate for arbitrary map implementations (e.g., tree-based, list-based, hash-based, etc.). 

A resource specification is \emph{valid} if all its actions commute modulo its abstraction function (property (4)), and every action's precondition suffices to ensure the low-ness of the abstract view of the data (property (3\peter{b})); a formal definition of validity will follow in \secref{logic}. The proof of validity has to be done only once per resource specification, and can be reused for different programs.

\peter{Crucially,} our commutativity criterion (property (4)) enforces (abstract) \emph{pairwise} commutativity of the actions in the resource specification, \peter{which avoids the (huge) effort of enumerating and comparing all possible interleavings of actions in a program:}
\marcon{Pairwise commutativity of all actions is sufficient to permute the schedule of one execution into any possible schedule of a second execution with the same results (modulo abstraction). }

\subsection{Program Verification}
\label{sec:bijection}

Given a program and a resource specification that satisfies  properties (3\peter{b}) and (4), it remains to check properties (1), \peter{(2), and (3a)} to prove that the program adheres to its resource specification. 

Our logic for doing so, \logicName, is a concurrent separation logic (CSL)~\cite{csl}.
To enable reasoning about low and high values, we phrase \logicName as a relational logic that proves properties of two executions of the same program\footnote{While our technique could in principle also reason about more than two executions at a time, we focus only on pairs of executions, since that is sufficient for proving non-interference.}.
 Like other concurrent separation logics, it requires showing that programs are data race free, by proving that each thread operates on its own partial heap separate from those of all other threads. 
 
\paragraph{Invariants} \peter{As is standard in separation logics, heap locations that may be modified by different threads} must be part of a shared resource, which is associated with an invariant that describes and constrains the resource's partial heap. \peter{In addition,} we use the invariant to also map the shared data to some pure value, as described before; that is, the invariant is used to connect a heap-based shared data structure in the verified program to a value of the  type defined in the resource specification. We denote the invariant as $I(v)$, where $v$ denotes the pure value the invariant maps its heap to. As explained before, for a linked list, $I(v)$ would typically be some predicate $\mathit{list}(p, v)$ for some pointer $p$ and sequence $v$. 
When initially sharing the resource, the invariant $I(v)$ must be established for some $v$. We check property (1) by enforcing that at this point, $\alpha(v)$ is low. 
\marcon{The invariant itself must not express any low-ness constraint.}

\paragraph{Atomic modifications} \peter{As explained in \secref{resource-specification}, we check property~(4) on the level of abstract actions rather than concrete implementations. For this to be sound, \emph{all} modifications of a shared data structure must be reflected by one of its actions, which we verify as follows.}

In \logicName (as in normal CSL), threads may modify the shared resource only in \code{atomic}-blocks. When entering such a block, they obtain the invariant, and on leaving the block, they have to re-establish the invariant and give it up again\footnote{\emph{Obtaining} the invariant adds the partial heap described by the invariant to the current heap; \emph{giving it up} amounts to checking that the current heap can be split into a partial heap that satisfies the invariant and a remainder, and removing the former. These logical steps are sometimes called \emph{producing} and \emph{consuming}, or \emph{inhaling} and \emph{exhaling} the invariant.}. 
We use this mechanism to check that a program modifies a shared data structure only via the legal actions of the associated resource specification. Since a shared data structure may be modified only within an \code{atomic}-block, we can impose the following proof obligation: If, at the beginning of an \code{atomic}-block, $I(v)$ holds for some $v$, then at its end, $I(v')$ holds for some $v'$ that is the result of applying one of the legal actions to the old value $v$, i.e., $v' = f_a(v, \mathit{arg})$ for some action $a$ and argument $\mathit{arg}$.

\paragraph{Guards}
\peter{Let's turn to the remaining two properties, (2) and (3a). We could enforce property~(2)} by proving that there are no \code{atomic}-blocks under high guards (i.e., inside conditionals or loops whose conditions are high). \peter{However, this check would be overly conservative because it rules out implementations where the final value of a shared data structure is low even though intermediate states depend on a secret; our evaluation demonstrates that such examples occur in practice.
}

\peter{Therefore, we follow an alternative approach: we \emph{record} \marcon{all actions that are performed on the shared resource}, and then check property~(2) retroactively on the recorded actions when the resource is unshared (and all concurrent modifications must therefore have finished). We will follow the same approach for property~(3a) and, thus, also record the arguments of each performed action, such that we can check action preconditions later.}

\peter{For each action, we introduce a separation logic resource, which we call \marco{a} \emph{guard}, to record how often and with which arguments this action has been performed. Like other separation logic resources, guards can be transferred between methods, split into fractional parts such that they can be shared between threads, and subsequently be recombined. Guards are parameterized with a multiset of arguments,
which records the arguments of the actions performed on a shared resource so far. \thibault{These parameters reflect} \emph{which} actions have been performed, \marcon{how often, and with which arguments, but not their order,} which is not known due to the influence of scheduling.} 

\peter{When sharing a resource, we obtain a guard for each action that is legal according to the resource specification. The parameters of these guards are initially empty because no actions have been performed yet. Performing an action then imposes a proof obligation that some fraction of the respective guard is held, and adds the argument to the guard's multiset.}

\peter{Guards allow us to check properties~(2) and~(3a) at the time when a resource is unshared. \thibault{Unsharing} requires all guards to be held, so that all performed actions are known.} Unsharing then consumes the guards, so that no threads have the ability to perform any actions after this point.
We show property~(2) \peter{simply} by proving that the \emph{cardinality} of the argument multiset (i.e., how often each action has been performed) is low.

\peter{To ensure property~(3a), we need to prove that each execution of an action satisfies its precondition. Since these preconditions can be relational (e.g., requiring an argument to be low), this proof must match the execution of an action in one run of the program with an execution in the other run. Proving preconditions retroactively when unsharing a resource gives the proof more freedom which executions of an action to match, which makes the proof technique more complete, and is possible because preconditions are expressed only over argument values and therefore state-independent.}

\peter{Concretely, when unsharing, we require showing for each action $a$ that there is a bijection between the elements of the multiset of arguments $s$ in one program execution and its elements in the other execution \marcon{(ensuring that their number is the same, i.e., the action has been performed the name number of times in both executions)}, such that each pair of elements  fulfills the relational precondition of the action. We denote this fact via the assertion $\allpre{a}{s}$.
}
For the \textsc{Put} action of the map example%
, this means checking that there is a bijection that maps every key-value pair in the multiset in the first execution to a pair with the same key (but potentially a different value) in the second execution, since the precondition of \textsc{Put} requires the keys to be low (but not the values).

\begin{figure}[t!]
\begin{minipage}[t]{0.48\textwidth}
\begin{lstlisting}[numbers=left]
procedure targets(households) {
  // ensures @[$\exists v' \ldotp \listpred{res}{v'} * \below{v'}{\llow}$ @]
  n := |households|
  m := createMap()
  @[\textcolor{blue}{$\{ \mappred{m}{\emptymap} \}$}@]
  @[\textcolor{blue}{$\Rightarrow \{ \mappred{m}{\emptymap} *  \below{\fdomain{\emptymap}}{\llow} \}$}@]
    // share
    @[\textcolor{blue}{\{ \aguard{\textsc{Put}}{\emptymultiset}{1} \}}@]
    @[\textcolor{blue}{$\Rightarrow \{ \aguard{\textsc{Put}}{\emptymultiset}{\frac{1}{2}} * \aguard{\textsc{Put}}{\emptymultiset}{\frac{1}{2}} \} $}@]
    worker(households, 0, n/2, m) || 
      worker(households, n/2, n, m)
    @[\textcolor{blue}{\{ $\exists s_1, s_2 \ldotp \aguard{\textsc{Put}}{s_1}{\frac{1}{2}} * \allpre{\textsc{Put}}{s_1} * $}@]
     @[\textcolor{blue}{  $\aguard{\textsc{Put}}{s_2}{\frac{1}{2}} * \allpre{\textsc{Put}}{s_2} \}$}@]
    @[\textcolor{blue}{$\Rightarrow \{ \exists s \ldotp \aguard{\textsc{Put}}{s}{1} * \allpre{\textsc{Put}}{s} \}$}@]
    // unshare
  @[\textcolor{blue}{$\{ \exists v \ldotp \mappred{m}{v} * \below{\fdomain{v}}{\llow} \}$}@]
  return sort(toList(keys(m)))
  @[\textcolor{blue}{$\{ \exists v, v' \ldotp \mappred{m}{v} * \below{\fdomain{v}}{\llow} * $}@]
   @[\textcolor{blue}{$\listpred{res}{v'} * \below{v'}{\llow} \}$}@]
}
\end{lstlisting}
\end{minipage}
\begin{minipage}[t]{0.42\textwidth}
\begin{lstlisting}[numbers=left]
procedure worker(households, f, t, m) {
  // requires @[$\aguard{\textsc{Put}}{\emptymultiset}{\frac{1}{2}} $@]
  // ensures @[$\exists s' \ldotp \aguard{\textsc{Put}}{s'}{\frac{1}{2}} * \allpre{\textsc{Put}}{s'} $@]
  @[\textcolor{blue}{$\{ \aguard{\textsc{Put}}{\emptymultiset}{\frac{1}{2}} \} $} @] 
  @[\textcolor{blue}{$\Rightarrow \{ \aguard{\textsc{Put}}{\emptymultiset}{\frac{1}{2}} * \allpre{\textsc{PUT}}{\emptymultiset} \} $} @] 
  for (i in f..t-1) {
    @[\textcolor{blue}{$\{ \exists s' \ldotp \aguard{\textsc{Put}}{s'}{\frac{1}{2}} * \allpre{\textsc{PUT}}{s'} \} $} @]
    adr, rsn := select(household[i])
    @[\textcolor{blue}{$\{ \exists s' \ldotp \aguard{\textsc{Put}}{s'}{\frac{1}{2}} * \allpre{\textsc{PUT}}{s'} * \below{adr}{\llow} \} $} @]
    atomic:
      @[\textcolor{blue}{$\{ \exists v \ldotp \mappred{m}{v} \} $} @]
      m.put(adr, rsn)
      @[\textcolor{blue}{$\{ \exists v \ldotp \mappred{m}{v[adr \mapsto rsn]} \} $} @]
      @[\textcolor{blue}{$\Rightarrow \{ \mappred{m}{f_{\textsc{Put}}(v, \langle adr, rsn \rangle)} \} $} @]
    @[\textcolor{blue}{$\{ \exists s' \ldotp \aguard{\textsc{Put}}{\msunion{s'}{\msliteral{\langle adr, rsn \rangle}}}{\frac{1}{2}} * $} @]
     @[\textcolor{blue}{$ \allpre{\textsc{Put}}{s'} * \lockpre{\textsc{Put}}{\langle adr, rsn \rangle} \} $} @] 
    @[\textcolor{blue}{$\Rightarrow \{ \exists s' \ldotp \aguard{\textsc{Put}}{\msunion{s'}{\msliteral{\langle adr, rsn \rangle}}}{\frac{1}{2}} * $} @]
     @[\textcolor{blue}{$ \allpre{\textsc{Put}}{\msunion{s'}{\msliteral{\langle adr, rsn \rangle}}} \} $} @]  
  }
}
\end{lstlisting}
\end{minipage}
\vspace{-3mm}
\caption{Proof outline for the example from Fig.~\ref{fig:mapkeyexample}, verified against the resource specification from Fig.~\ref{fig:examplelockspecmerged} (left). $I(v)$ is defined to be $\mappred{m}{v}$, where $\mathit{Map}$ is assumed to be a pre-existing separation logic predicate that \peter{relates the contents of a map $m$ to a mathematical value $v$. Superscript $\#$ denotes multiset operations.
$\aguard{\textsc{Put}}{s}{f}$ denotes an $f$-fraction of the guard for action \textsc{Put} with argument multiset $s$.}
} \label{fig:mapkeyexampleproof}
\end{figure}

\subsection{Verification of the Map Example}

We now show the entire proof for the map example from Fig.~\ref{fig:mapkeyexample} in Fig.~\ref{fig:mapkeyexampleproof}  (where we assume that our resource specification from Fig.~\ref{fig:examplelockspecmerged} (left) is valid). 
The code indentation in \code{targets} \peter{indicates where} the resource is shared. When sharing the resource, we must establish and give up the invariant $I(v)$, which we instantiate to $\mappred{m}{v}$ (a separation logic predicate for a map, which \peter{relates the map's contents to} the pure partial mapping $v$), and show that $\alpha(v)$ is low (property (1)), which is the case because the map is empty in both executions. We then share the map, meaning that we \peter{obtain} the guard for the \textsc{Put} action (the only action in our resource specification), with an empty argument multiset. We split this guard into two parts and give each part to a worker. 

Inside \code{worker}, we prove the loop invariant $\aguard{\textsc{Put}}{s'}{\frac{1}{2}} * \allpre{\textsc{Put}}{s'}$ for some $s'$ (the argument multisets in proofs are typically existentially quantified, s.t.\ it is not necessary to track actual values in specifications). That is, we collect all arguments with which we performed the \textsc{Put} action in $s'$ \marcon{(which also tracks how often we have performed the action)} and show that each such argument satisfies \textsc{Put}'s precondition. As explained above, we are  \emph{required} to prove $\allpre{\textsc{Put}}{s'}$ only later when we unshare the resource, but in this example, it is convenient to maintain this property throughout.
Initially, $\allpre{\textsc{Put}}{s'}$ holds trivially since $s'$ is empty.   

When the worker enters an \code{atomic}-block, it \peter{obtains} the invariant $I(v)$ for some $v$ \peter{(that is, $\mappred{m}{v}$)}, and has to show at the end of the block that $I(v')$ holds, where $v' = f_{\textsc{Put}}(v, \mathit{arg})$ and $\mathit{arg} = \langle \code{adr}, \code{rsn} \rangle$. 
We assume that the specification of the \code{put}-method is sufficiently strong to prove this. 
\marcon{Note that, since inside the \code{atomic}-block only the invariant (which cannot contain any low-ness constraints) is known about the shared data, if the code were to read shared data inside the block, the data would be implicitly treated as high.}
\peter{The proof rule for \code{atomic}-blocks adds} $\mathit{arg}$ to the argument multiset of the guard. Since we can prove at this point that $\mathit{pre}_{\textsc{Put}}(\mathit{arg})$ holds (i.e., \code{adr} is low), and $\allpre{\textsc{Put}}{s'}$ held for the old argument multiset $s'$, we can show that the new argument multiset $s'' =\msunion{s'}{\msliteral{\mathit{arg}}}$ also fulfills $\allpre{\textsc{Put}}{s''}$, and thus we maintain our loop invariant (which is also the postcondition of \code{worker}).

Back in \code{targets}, we unshare the resource, which consumes the guard. To do so, we recombine the fractional guards into a single one, whose argument multiset $s$ is the union of the argument multisets of the individual guards, $s_1$ and $s_2$. Now we \emph{must} prove that $\allpre{\textsc{Put}}{s}$ holds, \peter{which follows from the postcondition of the workers, once the guards are combined.} We may \emph{now} assume that the abstract view $\alpha(v)$ of the \marcon{\emph{current}} value $v$ of the shared map is low, i.e., we know that the map's key set is low and have to prove the postcondition of \code{targets}, \marcon{which states that the contents of the returned list are entirely low}. 
\marcon{We prove this using appropriate specifications of the procedures \code{keys}, \code{toList}, and \code{sort} that are called in the last line. For example, we use a postcondition for \code{keys} stating that if the key set of its input map is low, then the contents of the set it returns are low.}

\subsection{Unique Actions}

The technique we presented so far requires that \emph{all} modifications of the shared resource commute (modulo abstraction). However, in asymmetric thread collaborations, there are often actions that are performed by only one thread (e.g., in a producer-consumer scenario, a single producer thread might add values to a queue, whereas multiple consumer threads read and remove values from the queue); we call such actions \emph{unique} actions (as opposed to \emph{shared} actions that are performed by multiple threads). Since the order of applications of a unique action does not depend on scheduling, it is not necessary for unique actions to commute with themselves.

As an example, consider a variation of the map example where different workers work on different ranges of keys, and where both the keys and values \textsc{Put} into the map are low. In such a program, it is never the case that a \textsc{Put} of one thread overwrites a value that was \textsc{Put} by a different thread, and therefore differences in timing do not lead to different final map contents. 

We can support \peter{such examples} by defining an alternative resource specification that declares $n$ different \textsc{Put} actions, where the precondition of each action $\textsc{Put}_i$  requires its argument key to be in a range separate from that of all other \textsc{Put} actions (and both key and value to be low), as shown in Fig.~\ref{fig:examplelockspecmerged} (right) for $n=2$. We declare each of these actions to be unique, which means that they may  be performed only by a single thread. 
For resource specification validity, unique actions are required to (abstractly) commute with all actions \emph{except} themselves. That is, for example, $\textsc{Put}_1$ has to commute with $\textsc{Put}_2$, but \emph{not} with itself. 
We can now define $\alpha$ to be the identity function \peter{(i.e.,
use no abstraction at all), and prove programs secure even if they leak the entire map}\footnote{As an aside, this example also shows that the same data structure can have different resource specifications that have different demands and give different guarantees for the sensitivity of the final result.}.

During program verification, we do not allow splitting the guards for unique actions, which ensures that only a single thread may perform them. In addition, guards for unique actions record the \emph{sequence} of previous arguments, since their overall order is now no longer dependent on scheduling, and $\allpre{a}{s}$ for a unique action $a$ and argument sequence $s$ requires that the length of $s$ is low, and for each index $i$, the \marco{values} of $s[i]$ in both executions fulfill the action's precondition.

\section{Logic}\label{sec:logic}
In this section, we formalize our technique in \logicName, a relational concurrent separation logic with support for commutativity-based reasoning. 

\subsection{Language}
We formalize our logic for an imperative, concurrent language with a mutable heap, whose commands are defined in Fig.~\ref{fig:language}. 
We assume as given some expression language for integer-typed expressions $e$ and boolean expressions $b$, with the usual operations.
Commands like assignments, conditionals, loops, parallel composition and sequential composition are standard; \marcon{parallel compositions can be nested to create programs with more than two concurrent threads}. 
Programs interact with the heap via read and store commands $\cread{x}{e}$ and $\cwrite{e}{e}$, as well as $\alloc{x}{e}$, which allocates a single heap location, initializes its value to $e$, and assigns the resulting pointer to $x$; an extension to allocating multiple heap locations at once is straightforward.
\peter{Like other CSLs, we formalize our logic for} a single shared resource; threads can use the command 
$\atomic{c}$ to atomically execute the command $c$ while having access to the shared resource.  
However, our approach is not limited to having a single shared resource, and multiple resources are supported in our implementation analogous to other CSLs~\cite{csl,brookes}.

\begin{figure}[t]
\[
\begin{array}{lrll}
&c&{::=}&\assign{x}{e} 
 \mid \cread{x}{e}
 \mid \cwrite{e}{e} 
 \mid \alloc{x}{e} \mid \noop
 \mid \seq{c}{c}
   \mid \cond{b}{c}{c} \\
   &&&
 \mid \while{b}{c} 
  \mid \parc{c}{c}
 \mid \atomic{c}
\end{array}
\]
\vspace{-3mm}
\caption{Programming language. $e$ ranges over integer-typed expressions, $b$ over boolean-typed ones.}
\label{fig:language}
\end{figure}

Program states have the form $\langle \store, \nheap \rangle$, where the store $\store$ is a map from names to (integer) values and the heap $\nheap$ is a partial map from locations (natural numbers) to integer values. 
Expression evaluation is deterministic and total, meaning in particular that references to uninitialized variables evaluate to some default value; we denote the value of expression $e$ in store $\store$ by $\mmeval{e}{s}$.
Program configurations have the form $c, \langle \store, \nheap \rangle$ or $\abort$; the latter represents a failed computation (which \peter{will} never be reachable for verified programs).  
We use a small-step semantics \peter{with} transitions \peter{of} the form $c, \langle \store, \nheap \rangle \rightarrow c', \langle \store', \nheap' \rangle$ (for non-aborting steps).
Our semantics is identical to that of other concurrent separation logics~\cite{cslsound}; its rules are shown in \appendixref{opsem}. 

\subsection{Resource Specifications}
To simplify the formalization, we assume that our resource specification allows arbitrarily many unique actions but only \emph{one} shared action, which is not restrictive, since one can merge multiple shared actions into one s.t.\ the argument selects which action to perform.
Given this assumption, a resource specification for a resource value of type $T$ has the form $\langle \alpha, f_{a_s}, F_{a_u} \rangle$, where $\alpha$ is the abstraction function of type $T \rightarrow T_{\alpha}$ (for some \peter{mathematical} type $T_{\alpha}$), $f_{a_s}$ is the shared action, whose precondition we denote as $\mathit{pre}_{a_s}$, and $F_{a_u}$ is a family of unique actions, indexed by a finite type \thibault{$I$}; we will write $f_{a_i}$ as a synonym for $F_{a_u}(i)$ for all \thibault{$i \in I$}, and denote the precondition of unique action $f_{a_i}$ by $\mathit{pre}_{a_{i}}$. 
For each action $a$, $f_a$ is a function of type $T \rightarrow T_{\mathit{arg}_a} \rightarrow T$, where $T_{\mathit{arg}_a}$ is the argument type of the function.
Both preconditions are \emph{relational} preconditions on the argument of the action, i.e., they can require that (aspects of) the arguments are low. 
Thus, they are boolean-typed functions that take two arguments of type $T_{\mathit{arg}_a}$, where the two arguments denote the values of the argument in the two executions.
Note that the preconditions \emph{cannot}  constrain the resource value itself; \peter{we explain this limitation and show how to work around it} in \appendixref{partial}.

A resource specification is \emph{valid} iff (A) every action's relational precondition is sufficient to preserve the low-ness of the abstract view of the resource value (property (3\peter{b}) from Sec.~\ref{sec:overview}), and (B) all \emph{relevant} pairs of actions commute w.r.t.\ the abstract view (property (4) from Sec.~\ref{sec:overview}). Relevant pairs of execution are the shared action paired with all actions including itself, and every unique action paired with all unique actions \emph{except} itself; this reflects the fact that unique actions do not have to commute with themselves.
Formally, we define validity as follows:
\begin{definition}
A resource specification $\langle \alpha, f_{a_s}, f_{a_u} \rangle$ is valid iff
\begin{enumerate}[(A)]
  \item For all actions $a$, values $v$\thibault{, $v'$,} and arguments $\mathit{arg}, \mathit{arg}'$, if $\alpha(v) = \alpha(v')$ and $\mathit{pre}_a(\mathit{arg}, \mathit{arg}')$, then $\alpha(f_{a}(v, \mathit{arg})) = \alpha(f_a(v', \mathit{arg}'))$.
  \item For all pairs of actions $a, a'$ in $\{(a_s, a_s)\} \cup \{(a_s, a_i) | i \in I \} \cup \{(a_i, a_j) | i, j \in I \wedge i \neq j\}$ and all arguments $\mathit{arg}, \mathit{arg}'$, if $\alpha(v) = \alpha(v')$ then $\alpha(f_{a'}(f_a(v, \mathit{arg}), \mathit{arg}')) = \alpha(f_{a}(f_{a'}(v', \mathit{arg}'), \mathit{arg}))$.
\end{enumerate}
\end{definition}

\subsection{Extended Heaps}
While our semantics works on ordinary program heaps $\nheap$, we define our assertions and our logic on \peter{\emph{extended heaps}, 
an enriched notion of heaps that} represent both fractional permissions~\cite{boyland_permissions} and guards. 
An extended heap $\gheap$ is a triple $\langle \pheap, \gsheap, \guheaps \rangle$. $\pheap$ is a standard \emph{permission heap}~\cite{DBLP:conf/popl/BornatCOP05,cslsound} that can express \emph{partial ownership} of a heap location, that is, a
partial map from locations to pairs $\langle r, v \rangle$ of positive rational numbers of at most 1 and values $v$. For example, a permission heap that maps location $l$ to $\langle v, \frac{1}{2} \rangle$ denotes a half permission to the heap location $l$, where value $v$ is stored. 
Partial ownership allows multiple threads to concurrently \emph{read} the same heap location, since reading a location requires only some positive permission amount, whereas modifying its value requires a permission of 1. 

The shared action guard state $\gsheap$ and the family of unique action guard states $\guheaps$, where $\guheaps(i) = \guheap_i$, are specific to our technique. 
A guard for an action represents the \emph{right} to perform that action, and it tracks the arguments with which said action has already been performed. 

Every \emph{unique} guard state $\guheap_i$ is either $\bot$ or a \emph{sequence} of argument values of the unique action $a_i$, and represents (when a resource is shared) the entire sequence of arguments with which the unique action has been performed. 
Since only one thread is allowed to execute a unique action, \peter{the order of executions is known and, thus, we can track the arguments in a sequence (as opposed to a multiset for shared actions). Moreover,} unique guard states cannot be split or combined; that is, the thread that performs the unique action will have the entire sequence of argument values in its unique guard state, and all other threads will have a guard state of $\bot$. Thus, when adding two unique guard states, if one has a non-$\bot$ value, then the other must be $\bot$, otherwise addition is undefined (see \appendixref{logicdefs} for the formal definition of guard heap addition).

Addition of guard heap families is defined pointwise. We write $\bot$ for guard heap families that are $\bot$ for all $i$, and $\guheapsingle{\guheap}{i}$ for a guard heap family whose value is $\guheap$ for $i$, and $\bot$ for all other indices.

The \emph{shared} guard state $\gsheap$ is either $\bot$ or a pair $\langle r, \mathit{args}_s \rangle$, where $r$ is a positive rational number of at most 1 and $\mathit{args}_s$ is a \emph{multiset} of argument values.
The shared action may be performed by multiple threads if each of those threads has a positive fractional shared guard state. Thus, the shared guard state represents the (potentially partial) knowledge of the multiset of arguments with which the shared action has been executed so far. 
If $r$ is $1$, then $\mathit{args}_s$ denotes \emph{all} arguments with which the action has been performed so far. 
On the other hand, two threads might have  guard states with $r=\frac{1}{2}$ each and respective arguments $\mathit{args}_1$ and $\mathit{args}_2$, which each contain all arguments with which \emph{that thread} has performed the shared action; then, in total, the action has been performed with the union of the multisets $\mathit{args}_1 \msadd \mathit{args}_2$. In general, when adding the arguments of two non-$\bot$ shared guard states, we take the union of the argument multisets, or that of one state if the other is $\bot$.
The sum of two permission heaps $\pheap \oplus \pheap'$ is standard; intuitively, permission amounts are added (to a value of at most one) and values are unchanged (see \appendixref{logicdefs}).

The sum of two extended heaps $\langle \pheap, \gsheap, \guheap \rangle \oplus \langle \pheap', \gsheap', \guheap' \rangle$ is defined as $\langle \pheap \oplus \pheap', \gsheap \oplus \gsheap', \guheap \oplus \guheap' \rangle$ iff the sums of all its components are defined.
Finally, an extended heap can be \emph{normalized} to a normal heap (which we denote by $\normalize{\gheap}$), by taking the permission heap and removing the permission amounts. That is, the normalized form of $\langle \pheap, \gsheap, \guheap \rangle$ has the domain $\funcdom{\pheap}$, and for each location $l \in \funcdom{\pheap}$, the normalized heap has the value $v$ s.t.\ $\pheap(l) = \langle \_, v \rangle$.

\subsection{Assertions}

Our assertion language is \thibault{defined as follows:
$$P,Q{::=} \mathit{emp} \mid b \mid \pointsto{e}{r}{e} \mid P \ast Q \mid  P \wedge Q 
\mid \exists x \ldotp P \mid
\aguardlbl{a}{r}{}{e} \mid \aguardnolbl{u}{i}{e} 
\mid b \Rightarrow P \mid \below{e}{\llow}
$$
where $P$ and $Q$ range over assertions,
$r$ over positive rationals up to 1,
$b$ ranges over boolean
and $e$ over all expressions (including sequence- and multiset-typed expression,
to describe the arguments of the two kinds of guard-assertions).}
Assertions are \emph{relational}: their validity (Fig.~\ref{fig:assertion-validity}) is defined over pairs of states, allowing them to express that expressions are \peter{low, i.e.,} equal in both states.

\begin{figure}
{\footnotesize
\[
\begin{array}{rlcl}
(\store_1, \gheap_1), (\store_2, \gheap_2) \models & \mathit{emp} &\Longleftrightarrow& \funcdom{\pheap_1} = \emptyset \wedge \funcdom{\pheap_2} = \emptyset \\
(\store_1, \gheap_1), (\store_2, \gheap_2) \models & b &\Longleftrightarrow& \expval{b}{\store_1} = 1 \wedge \expval{b}{\store_2} = 1 \\
(\store_1, \gheap_1), (\store_2, \gheap_2) \models  & \pointsto{e_1}{r}{e_2} &\Longleftrightarrow& 
\gheap_1 = \{\expval{e_1}{\store_1} \mapsto \langle r, \expval{e_2}{\store_1}\rangle \} \wedge  \gheap_2 = \{\expval{e_1}{\store_2} \mapsto \langle r, \expval{e_2}{\store_2}\rangle \}\\
(\store_1, \gheap_1), (\store_2, \gheap_2) \models & P \ast Q &\Longleftrightarrow& \exists \gheap'_1, \gheap''_1, \gheap'_2, \gheap''_2 \ldotp 
\gheap_1 = \gheap'_1 \oplus \gheap''_1 \wedge \gheap_2 = \gheap'_2 \oplus \gheap''_2 \wedge  \\
& & & (\store_1, \gheap'_1), (\store_2, \gheap'_2) \models P \wedge (\store_1, \gheap''_1), (\store_2, \gheap''_2) \models Q \\
(\store_1, \gheap_1), (\store_2, \gheap_2) \models & P \wedge Q &\Longleftrightarrow&  (\store_1, \gheap_1), (\store_2, \gheap_2) \models P \wedge (\store_1, \gheap_1), (\store_2, \gheap_2) \models Q \\
(\store_1, \gheap_1), (\store_2, \gheap_2) \models & \exists x \ldotp P &\Longleftrightarrow& \exists v_1, v_2 \ldotp (\store_1[x \mapsto v_1], \gheap_1), (\store_2[x \mapsto v_2], \gheap_2) \models P \\
(\store_1, \gheap_1), (\store_2, \gheap_2) \models & \aguardlbl{s}{r}{}{e} &\Longleftrightarrow& \gsheap_1 = \langle r, \expval{e}{\store_1} \rangle \wedge \gsheap_2 = \langle r, \expval{e}{\store_2} \rangle \wedge  \\ & & & \guheaps_1 = \bot \wedge \guheaps_2 = \bot \wedge \funcdom{\pheap_1} = \emptyset \wedge \funcdom{\pheap_2} = \emptyset \\
(\store_1, \gheap_1), (\store_2, \gheap_2) \models & \aguardnolbl{u}{i}{e} &\Longleftrightarrow& \guheaps_1 = \guheapsingle{\expval{e}{\store_1}}{i} \wedge \guheaps_2 = \guheapsingle{\expval{e}{\store_2}}{i} \wedge  \\ & & & \gsheap_1 = \bot \wedge \gsheap_2 = \bot \wedge \funcdom{\pheap_1} = \emptyset \wedge \funcdom{\pheap_2} = \emptyset \\
(\store_1, \gheap_1), (\store_2, \gheap_2) \models & b \Rightarrow P  &\Longleftrightarrow& \expval{b}{\store_1} = \expval{b}{\store_2} \wedge \expval{b}{\store_1} \Rightarrow (\store_1, \gheap_1), (\store_2, \gheap_2) \models P \\
(\store_1, \gheap_1), (\store_2, \gheap_2) \models & \below{e}{\llow}  &\Longleftrightarrow& \expval{e}{\store_1} = \expval{e}{\store_2}
\end{array}
\]
}%
\vspace{-3mm}
\caption{Assertion validity, where $\gheap_i = \langle \pheap_i, \gsheap_i, \guheaps_i \rangle$. 
}
\label{fig:assertion-validity}
\end{figure}

All basic separation logic assertions have their usual meaning, but applied to both states. \peter{For instance, standard points-to-assertions $\pointsto{e_1}{r}{e_2}$ represent a permission amount of $r$ to location $e_1$ with a value of $e_2$.} Separating conjunctions $P \ast Q$ hold if the extended heaps of \emph{both} states can be separated into partial extended heaps that fulfill the conjuncts. 
Existentials are interpreted such that there can be \emph{different} values for the quantified variable in both states, \peter{that is, 
$\exists x \ldotp \pointsto{e}{1}{x}$ expresses that $e$ points to potentially different values in the two} states (i.e., it might be high).

The assertion $\aguardlbl{s}{r}{}{e_a}$ represents a fractional amount $r$ of the guard for the shared action with the multiset of  arguments $e_a$; it holds \thibault{in a state with store $s$} iff the shared guard state is \thibault{$\langle r, \expval{e_a}{s} \rangle$}, the permission heap is empty, and the unique guard state is $\bot$. Thus, a state fulfilling $\aguardlbl{s}{r}{}{e_a}$ can be added to any other state with a compatible shared guard state. 
Similarly,  $\aguardnolbl{u}{i}{e_a}$ represents the guard for the unique action with index $i$ and holds iff the sequence of arguments is \thibault{$\expval{e_a}{s}$}, the shared guard state is $\bot$, and the permission heap is empty.

Finally, the relational assertion $\below{e}{\llow}$ states that expression $e$ is low. When used in an implication $b \Rightarrow \below{e}{\llow}$, it can also express \emph{value-dependent sensitivity}~\cite{covern}. 
For example, a data structure might contain pairs of booleans and other values, where the boolean expresses the sensitivity of the other value; then, at runtime, clients could retrieve values from the data structure and e.g.\ output them to different channels based on their sensitivity.

We define $\allpre{}{e}$ in terms of other assertion constructs\footnote{\thibault{In our Isabelle formalization, $\allpre{}{e}$ is actually defined as another assertion in the assertion language.}}.
$\allpre{s}{e}$, for the shared action, expresses that there is a bijection between the elements of the  multiset-typed expression $e$ in one state and the elements of $e$ in the other state, s.t.\ each element in one state and its corresponding element in the other state together fulfill the relational precondition $\mathit{pre}_{a_s}$ of the shared action. Formally, we define this notion recursively, by stating that there is an element $x$ in $e$ in one execution, and an element $x'$ in $e$ in the second execution, s.t.\ they together fulfill $\mathit{pre}_{a_s}$, and after removing both values from their respective multisets, $\allpre{}{e}$ holds again for the resulting smaller multisets: 
\begin{definition}
If the argument type of the shared action is $T_{\mathit{arg}_s}$ and $e$ is a multiset of $T_{\mathit{arg}_s}$, then 
\begin{equation}
  \allpre{s}{e} = 
  \begin{cases}
    \trueval &\text{ if }e = \emptymultiset \\
    \exists x \in e \ldotp \lockpre{a_s}{x} \wedge \allpre{s}{e \msminus \msliteral{x}} & \text{otherwise} \\
  \end{cases}
\end{equation}
\peter{Since $\mathit{pre}_{a_s}$ is relational, $\allpre{s}{e}$ is also relational. Moreover, since the existential quantifier may be interpreted differently in both executions,  $\allpre{s}{e}$ expresses the existence of a bijection between the argument multiset, as discussed in \secref{bijection}.}

For each unique action, we require that for the sequence of arguments $e$, \marcon{the length of the sequence is low and} the elements at each index fulfill the action's relational precondition: 
\begin{equation}
  \allpre{i}{e} = \below{|e|}{\llow} \wedge
  \bigwedge_{j \in [0, |e|)}{\lockpre{a_i}{e[j]}}
\end{equation}
\end{definition}

We call an assertion \emph{unary} if it does not restrict pairs of state relative to one another. That is, $P$ is unary if for all $\store_1, \gheap_1, \store_2, \gheap_2$, if
$(\store_1, \gheap_1), (\store_1, \gheap_1) \models P$ and $(\store_2, \gheap_2), (\store_2, \gheap_2) \models P$ then also $(\store_1, \gheap_1), (\store_2, \gheap_2) \models P$. Note that all assertions that do not syntactically contain any $\below{e}{\llow}$-assertions (or $\allpre{}{e}$-assertions that are defined in terms of $\below{e}{\llow}$) are unary.

Finally, $\noguard{P}$ states that the assertion $P$ holds only for states whose guard states are all $\bot$.

\subsection{Resource Contexts}

A \emph{resource context} combines a resource specification with an assertion that defines the shared heap data structure and maps it to a single \peter{mathematical} value. 
That is, a resource context $\Gamma$ has the form $\langle \alpha, f_{a_s}, F_{a_u}, I(x) \rangle$, where the \emph{invariant} $I(x)$ is a parameterized unary assertion defining a valid state of the shared data structure with the value $x$. 
The value of $x$ must be uniquely defined by the invariant, i.e., we require that if $(\store_1, \gheap_1), (\store_2, \gheap_2) \models I(v)$ and $(\store_1, \gheap_1), (\store_2, \gheap_2) \models I(v')$ then $v = v'$. 

Given a \peter{resource} context $\Gamma$, we can now formally connect the value of the guard states to the permission heap, using the notion of \emph{consistency}. 
Consistency expresses that the value of the shared data structure on the heap is a possible result of applying the recorded actions with the recorded arguments in some order.
We say that a state with shared guard state $\langle r, \mathit{args}_s \rangle$ and unique guard states $\mathit{args}_i$ for every unique action $a_i$ is consistent with a context $\Gamma$ from an initial value $v_0$ iff
\thibault{(1)}~its shared guard state is complete  (i.e., $r = 1$),
\thibault{(2)} $I(v)$ holds in the state, for some $v$, and
  \thibault{(3)} if $FU_i$ is the sequence of applications of $f_{a_i}$ with the arguments in $\mathit{args}_i$ (i.e., the first application in $FU_i$ applies $f_{a_i}$ with argument $\mathit{args}_i[0]$, etc.) for all $i$, and similarly, $FS$ is the sequence of applications of $f_{a_s}$ with the arguments in \emph{some} permutation of $\mathit{args}_s$, then there is some interleaving of $FS$ and all $FU_i$ s.t.\ applying this interleaving to the initial value $v_0$ results in a final value of $v$.

\subsection{Proof Rules}

Our proof rules define a judgment of the form $\mmtriple{\Gamma_{\bot}}{P}{c}{Q}$, where $\Gamma_{\bot}$ is either $\bot$ or some resource context $\Gamma$. Intuitively, it expresses that a command $c$, when executed from two states that fulfill $P$ and either no initial shared resource (if $\Gamma_{\bot}$ is $\bot$) or a shared resource fulfilling $\Gamma$, will not abort, and if they terminate, the resulting states will fulfill $Q$. The formal definition will follow in \secref{soundness}. 

\figref{rules} shows the most important proof rules of \logicName, those for sharing \peter{and unsharing, as well as for} atomic actions. 
All other rules, e.g., those for assignments, \peter{heap accesses, as well as sequential and parallel composition, and the rules for framing, existentials, and consequence}, are standard and shown in \appendixref{logic}. The rules for loops and conditionals are similar to other relational logics \marcon{but, crucially, do not require conditions to be low, and thus allow secret-dependent branching. If branch conditions are high, then postconditions have to be unary, which prevents indirect flows by making it impossible to have $\below{e}{\llow}$ in the postcondition. 
}

The share rule \peter{captures both the share and unshare operation. It} can be used only when the current resource context is $\bot$, i.e., when there is no shared data structure. Once the share rule is used, threads can access a shared data structure, and must use guards to justify their actions on it and record the arguments of the actions they perform. The share rule enforces that the new context's resource specification is valid, the resource invariant initially holds, and the value $x$ of the resource is low modulo abstraction \peter{(property~(1) from \secref{abstract-commutativity})}. The postcondition of the conclusion expresses that the invariant will hold again, with a new value $x'$, whose abstract view is also low \marcon{(and thus allows assuming that the abstract view of the \emph{current} value of the shared data structure is low after concurrent modification has finished)}. In its premise, it gives access to the guards for the shared and the unique actions, initially with an empty argument multiset resp.\ argument sequences. It requires that after executing $c$, the guards are present again; in particular, the guard for the shared action must be present in its entirety, and therefore record all arguments of the shared action in its multiset. It also requires that 
$\mathit{PRE}$ holds for both the sequences of arguments of the unique actions and the multiset of arguments of the shared action, meaning that we can now, retroactively, show that for each application of an action, the relational precondition was fulfilled \peter{(property (3a))}.

There are two similar atomic-rules, one for the shared action and one for the unique actions. Both require that the resource context is not $\bot$, i.e., there is currently a shared resource, and that the guard for the respective action is initially present, which ensures that no \peter{thread can modify the shared data structure without holding} the appropriate guard. 
For the shared action, any positive fraction of the guard is sufficient; for the unique actions, the whole unsplittable guard is required. In the postcondition of the conclusion, the guard records the new argument with which the action has been performed in its multiset or sequence, respectively. The rule's premise requires showing that, assuming the invariant holds initially for some value $x_v$ \marcon{(which, since the invariant is unary and thus no low-ness assumptions can be made, forces one to treat all shared data as high)}, it holds again after the atomic modification, with a new value that is the result of applying the respective action with the respective argument. 
We impose side conditions that require the assertions $P$ and $Q$ to not contain any guards \peter{(i.e., any guards held before the atomic block must be framed away using the frame rule)}, and that the invariant $I(x)$ is precise~\cite{precise_predicates},
i.e., fulfilled by at most one sub-heap of any given heap. \peter{This condition is not limiting in practice, see \appendixref{sideconds}.}

\begin{figure*}
\[
\footnotesize
\begin{array}{c}
\Inf[\textsc{Share}]{\Gamma = \langle \alpha, f_{a_s}, f_{a_u}, I(x) \rangle}{\Gamma\text{ is valid}}{I(x)\text{ is unary and precise}}.
                    {\mmtriple{\Gamma}{P \ast \aguardlbl{s}{1}{}{\emptymultiset} \ast \mathit{UniqueEmpty}}{c}{Q \ast \aguardlbl{s}{1}{}{x_s} \ast \allpre{s}{x_s} \ast \mathit{UniquePre}}}
                    {\mmtriple{\bot}{I(x) \ast \below{\alpha(x)}{\llow} \ast P}{c}{\exists x' \ldotp I(x') \ast \below{\alpha(x')}{\llow} \ast Q}} \\
\ \\
\Inf[\textsc{AtomicShr}]{\Gamma = \langle \alpha, f_{a_s}, f_{a_u}, I(x) \rangle}{I(x)\text{ is unary and precise}}.
                        {x_v \notin \freevars{P} \cup \freevars{Q} \cup \freevars{I(x)}}{x_s, x_a, x_v \notin \modset{c}}{\noguard{P}}{\noguard{Q}}.
                   {\mmtriple{\bot}{P \ast I(x_v)}{c}{Q \ast I(f_{a_s}(x_v, x_a))}}
                   {\mmtriple{\Gamma}{P \ast \aguardlbl{a}{r}{}{x_s}}{\atomic{c}}{Q \ast \aguardlbl{a}{r}{}{\msunion{x_s}{\msliteral{x_a}}}}} \\
\ \\
\Inf[\textsc{AtomicUnq}]{\Gamma = \langle \alpha, f_{a_s}, f_{a_u}, I(x) \rangle}{I(x_v)\text{ is unary and precise}}.
                        {x_v \notin \freevars{P} \cup \freevars{Q} \cup \freevars{I(x)}}{x_s, x_a, x_v \notin \modset{c}}{\noguard{P}}{\noguard{Q}}.
                   {\mmtriple{\bot}{P \ast I(x_v)}{c}{Q \ast I(f_{a_i}(x_v, x_a))}}
                   {\mmtriple{\Gamma}{P \ast \aguardnolbl{a}{i}{x_s}}{\atomic{c}}{Q \ast \aguardnolbl{a}{i}{x_s ++ [x_a]}}} \\
\ \\
\end{array}
\]
\vspace{-3mm}
\caption{The central proof rules: the \textsc{Share} rule and the two \textsc{Atomic} rules implement all checks specific to our technique. 
\thibault{Recall that the family of unique actions $f_{a_u}$ is indexed by a (finite) type $I$;
we denote the elements of $I$ by $i_0, \ldots, i_n$.
We write $\mathit{UniqueEmpty}$ to abbreviate $\aguardnolbl{u}{i_0}{[]} \ast \dots \ast \aguardnolbl{u}{i_n}{[]}$,
and $\mathit{UniquePre}$ to abbreviate
$\exists x_{i_0}, \ldots, x_{i_n} \ldotp \aguardnolbl{u}{i_0}{x_{i_0}} \ast \allpre{u}{x_{i_0}} \ast \dots \ast \aguardnolbl{u}{i_0}{x_{i_n}} \ast \allpre{u}{x_{i_n}}$.
}
We use \thibault{$\freevars{P}$} to denote the free variables in \thibault{the assertion P},
and $\modset{c}$ for the set of variables modified by command $c$.}\label{fig:rules}
\end{figure*}

\subsection{Limitations}\label{sec:limitations}
\marcon{
\logicName is not complete and can be extended in multiple ways. Its central limitations are that
(1) it does not support information security proofs based on the absence of secret-dependent timing and thus cannot prove shared data low while it is being concurrently modified,
(2) for programs that enforce an ordering between concurrent modifications, requiring commutativity of \emph{all} actions is unnecessarily strong,
(3) in specific situations, it would be sound to treat intermediate values of shared data as low, which \logicName does not allow, and 
(4) it focuses on proving the output values of a program low and has no support for reasoning about the sensitivity of I/O (e.g., calls to \code{print}-procedures).
Limitation (1) can be mitigated by combining \logicName with existing logics like SecCSL and limitation (4) can be lifted via a simple extension of the logic (which we do in our implementation). Addressing limitations (2) and (3) is possible and left as future work.}

\marcon{In addition, \logicName does not change the fact that concurrent code handling secret data has to be specifically designed to be secure. While with existing techniques, code must avoid all secret-dependent timing, our technique provides a new pattern for designing secure code based on commutativity. It is well-suited for programs that concurrently compute and return data in the presence of secrets, but other techniques are more complete for programs that output values during concurrent modifications, if their assumptions about timing are fulfilled.}

\section{Soundness} \label{sec:soundness}
We have formalized \logicName and proved it sound in Isabelle/HOL; here, we give a high-level overview of our proof (which is  available in the Archive of Formal Proofs~\cite{FormalizationAFP}).
We build on Vafeiadis's soundness proof of CSL~\cite{cslsound}, 
whose basic idea is to define a predicate $\mathit{safe}_n(P, c, Q)$, which (ignoring frames and resources for now) expresses that $c$, when executed for $n$ steps from any state satisfying $P$, will not abort and, if it terminates \peter{after those $n$ steps}, will end up satisfying $Q$. Hoare triples are then defined to hold if $\mathit{safe}_n$ holds for all $n$. 
The same idea is used in the soundness proof for SecCSL~\cite{seccsl}, a concurrent separation logic for non-interference.
SecCSL 
enforces that two executions with identical low inputs have the same control flow, s.t.\ the scheduler makes the same decisions in both executions, and therefore (in an idealized scenario) no internal timing channels exist.
Since we allow high data to influence control flow and therefore timing and scheduling, 
we have to prove that, given two executions with identical low inputs with arbitrary, potentially \emph{different} schedules (and different control flow per thread), the program's public output will be low.

As a result, the inductive argument used in existing proofs does not work for our setting.
We solve this problem by separating the proof of safety from the proof of relational properties. 
We define a predicate $\mathit{safe}_n(\Gamma_{\bot}, c, \sigma, \Sigma)$, where $\Gamma_{\bot}$ is, as before, either $\bot$ or a resource context $\Gamma$, $\sigma$ ranges over program states with extended heaps of the form $\langle \store, \gheap \rangle$, and $\Sigma$ is a set of such program states. 
Intuitively, $\mathit{safe}_n$ expresses 
(1) that executing $c$ from $\sigma$, extended with a heap satisfying the resource invariant, for $n$ steps will not abort, 
(2) that if it terminates, will result in \emph{one} of the states in $\Sigma$ (since, due to non-deterministic scheduling, there may be more than one final state), 
and (3) that if there is a shared resource, the state remains consistent w.r.t.\ some initial value $v_0$, i.e., that the state of the resource heap continues to be a possible result of applying the actions with the arguments specified in the guards to the initial value $v_0$.
We provide the formal definition of safety in \appendixref{soundness}.
Crucially, this predicate makes a statement only about an individual execution.

We can now define the validity of a Hoare triple $\mmttriple{\Gamma_{\bot}}{P}{c}{Q}$ by stating that any two executions from a pair of states satisfying $P$ are $\mathit{safe}$ for an arbitrary number of steps, and \emph{any} pair of final states from those two executions will fulfill $Q$. More precisely, the Hoare triple holds if there is some function $\statefunc$ that maps any initial state $\pstate$ to a set of all possible final states $\Sigma$, and for any $\pstate_1, \pstate_2$ fulfilling $P$, all pairs of states in $\statefunc(\pstate_1) \times \statefunc(\pstate_2)$ fulfill $Q$: 
\begin{definition}
  $\mmttriple{\Gamma_{\bot}}{P}{c}{Q}$ holds iff there is some $\statefunc$ s.t.\ for all $n, \pstate$, if $\sigma, \sigma \models P$, we have $\mathit{safe}(n, \Gamma_{\bot}, c, \sigma, \statefunc(\pstate))$, and for all $\pstate_1, \pstate_2, \pstate'_1, \pstate'_2$, if $\pstate_1, \pstate_2 \models P$ and  $\pstate'_1 \in \statefunc(\pstate_1)$ and $\pstate'_2 \in \statefunc(\pstate'_2)$, then $\pstate'_1, \pstate'_2 \models Q$.
\end{definition}

\marcon{Our soundness proof uses a lemma stating, essentially, that the conditions (1)--(4) from Sec.~\ref{sec:overview} are sufficient to ensure the abstraction of the final shared value is low: 
}
\marcon{
\begin{lemma}
If $\Gamma$ is valid, $\alpha(v_0) = \alpha(v'_0)$, $v$ is consistent with $v_0$ through some sequence of shared action applications with argument multiset $\mathit{args}_s$ and unique action applications with argument sequences $\mathit{args}_i$ for all $i$, and similarly  $v'$ is consistent with $\Gamma$ from $v'_0$ through some sequence of shared action applications with argument multiset $\mathit{args}'_s$ and unique action applications with argument sequences $\mathit{args}'_i$ s.t.\ $\allpre{s}{\mathit{args}_s, \mathit{args}'_s}$ and $\allpre{i}{\mathit{args}_i, \mathit{args}'_i}$ for all $i$, then $\alpha(v) = \alpha(v')$.
\end{lemma}
}

\marcon{Here, condition (1) is $\alpha(v_0) = \alpha(v'_0)$, conditions (2) and (3a) are expressed by the  $\mathit{PRE}$-constraints, and the validity of $\Gamma$ represents conditions (3b) and (4). Using this lemma, we prove:}

\begin{theorem}\textnormal{(Soundness)}
If $\mmtriple{\Gamma_{\bot}}{P}{c}{Q}$ then $\mmttriple{\Gamma_{\bot}}{P}{c}{Q}$.
\end{theorem} 

\marcon{It follows that \emph{intermediate} assertions proved at some program point will actually hold at this point for any pair of executions that reaches it, since it is always possible to cut off the rest of the program (and the proof) after such a point and apply the soundness theorem to the part of interest.}

Thus, Hoare triples make true statements about pairs of program executions from an initial state with no shared resource (using $\gheapfg$ to range over extended heaps whose guard states are $\bot$ and whose permission heaps have full permission of every location in their domain):

\begin{corollary}
If $\mmtriple{\bot}{P}{c}{Q}$ and $\langle s_1, \gheapfg_1 \rangle, \langle s_2, \gheapfg_2 \rangle \models P$, then $c$ does not abort from any $\langle s_i, \normalize{\gheapfg_i} \rangle$, and if 
$c, \langle s_i, \normalize{\gheapfg_i} \rangle \rightarrow^* \noop, \langle s'_i, \nheap_i \rangle$ for some $s'_i, \nheap_i$ and $\nheap_i = \normalize{\gheapfg'_i}$ for some $\gheapfg'_i$, 
then $\langle s'_1, \gheapfg'_1 \rangle, \langle s'_2, \gheapfg'_2 \rangle \models Q$.
\end{corollary}

\marcon{Finally, we can prove non-interference according to Def.~\ref{def:ni} by proving that if all low input variables $I_l$ have low values initially, then all low output variables $O_l$ have low values in the end:}

\begin{corollary}
If $\mmtriple{\bot}{P \ast \bigwedge_{x \in I_l} \below{x}{\llow}}{c}{\bigwedge_{x \in O_l} \below{x}{\llow}}$ for some unary $P$ and we have $\langle s_1, \gheapfg_1 \rangle, \langle s_2, \gheapfg_2 \rangle \models P \ast \bigwedge_{x \in I_l} \below{x}{\llow}$, then if 
$c, \langle s_i, \normalize{\gheapfg_i} \rangle \rightarrow^* \noop, \langle s'_i, \nheap_i \rangle$ for some $s'_i, \nheap_i$ and $\nheap_i = \normalize{\gheapfg'_i}$ for some $\gheapfg'_i$, 
then $\langle s'_1, \gheapfg'_1 \rangle, \langle s'_2, \gheapfg'_2 \rangle \models \bigwedge_{x \in O_l} \below{x}{\llow}$.
\end{corollary}

\section{Implementation and Evaluation}\label{sec:impl}
We have implemented our technique in \toolName, an automated prototype verifier based on the Viper verification infrastructure~\cite{viper}, which is available as open source\footnote{\url{https://github.com/viperproject/hyperviper}}. 
\toolName supports a richer language than the one used in this paper; in particular, instead of parallel composition commands, it allows dynamic thread creation using \code{fork} and \code{join} commands. Additionally, \toolName supports \emph{multiple} resources in a single program (which can be associated with different resource specifications). 

\toolName's level of automation is similar to existing automated verifiers: Users must provide method pre- and postcondition as well as loop invariants, in an assertion language that is similar to the one shown in this paper, and additionally have to declare resource specifications. Given these specifications and some annotations indicating key proof steps (e.g., which action is performed by an atomic block, and how to split and merge shared action guards; see \appendixref{hyperviper} for an example), our tool automatically either verifies the program or indicates potential errors.

Internally, \toolName encodes the validity constraints for all resource specifications as well as all other proof obligations imposed by our logic into the Viper intermediate language. To encode relational proof obligations, it uses a modular product program construction~\cite{mpp}, and combines it with existing encodings for concurrent programs~\cite{peterrustan} in a sound way~\cite{cav21}.
Subsequently, it automatically verifies the generated program  using one of Viper's backend verifiers and, ultimately, the Z3 SMT solver~\cite{z3}. 

\begin{table}[]
\small
\begin{tabular}{l|l|l|r|r|r}
Example                 & Data structure          & Abstraction                   & LOC & Ann. & $T$ \\ \hhline{=|=|=|=|=|=}
Count-Vaccinated                 & Counter, increment               & None                          &  44 &  46  &  10.15   \\
\figref{setexample}             & Integer, add                      & None                          & 129 &  95  &  10.90   \\
Count-Sick-Days             & Integer, add                      & None                          & 52 &  45  &  13.67   \\
\figref{initexample}                & Integer, arbitrary                & Constant                      & 29 &  20   &  1.52   \\
Mean-Salary               & List, append                      & Mean                          & 80  &  84   &  14.10   \\
Email-Metadata           & List, append                      & Multiset                      & 82 &  75   &   16.70  \\
Patient-Statistic        & List, append                      & Length                          & 73 &  70    &   4.92  \\
Debt-Sum               & List, append                      & Sum                          & 76 &  81    &   14.45  \\
Sick-Employee-Names             & Treeset, add                        & None                          & 105  &  113  &   28.43  \\
Website-Visitor-IPs               & Listset, add                        & None                          & 74  &  69  &   6.20  \\
\figref{mapkeyexample}           & HashMap, put                          & Key set                        & 129  &  96   &   10.37  \\
Sales-By-Region             & HashMap, disjoint put              & None                          &  129 &  104  &   12.37  \\
Salary-Histogram                  & HashMap, increment value                       & None                          &  135 &  109   &  13.78   \\
Count-Purchases                  & HashMap, add value                       & None                          &  137 &  109   &  11.73   \\
Most-Valuable-Purchase        & HashMap, conditional put                     & None                          & 140  & 118    &   17.87  \\ \hline
1-Producer-1-Consumer   & Queue           & Consumed sequence             & 82 &  88   &  3.23   \\
Pipeline                & Two queues      & Consumed sequences      & 122 &  100    &    3.66 \\
2-Producers-2-Consumers & Queue         & Produced multiset             & 130  &  134   &   8.45  \\
\end{tabular}
\caption{Evaluated examples. We show the used data structure and the operation(s) we allow on it; for the last three examples, these are producing and consuming data. LOC are lines of code, not including specifications, Ann.\ are only specifications and proof annotations. $T$ is the verification time in seconds, averaged over 5 runs.}\label{tbl:eval}
\end{table}

To demonstrate the practical usefulness of our verification technique, 
we have applied \toolName to a number of example programs that \peter{represent a variety of applications that concurrently manipulate shared data structures with secret data}.
The first \peter{15} examples model specific applications, which spawn a number of worker threads to modify some shared data structure and subsequently output the abstract view of the shared data structure.
All examples explicitly model data structures containing different kinds of secret information (e.g., employee names and salaries, patient health data, or user activity on a website). Depending on the example, the data contains some low parts (e.g., whether or not a patient has been vaccinated), or some \emph{aspect} of the secret data is low (e.g., the range of an employee's salary or the number of---otherwise secret---purchases they have made). 
 
The last three examples model variations of \peter{general parallel programming patterns}; here, we leave the data and computation abstract and show instead that the pattern is safe for \emph{all} possible implementations where the individual roles do not directly leak secret data themselves.
Table~\ref{tbl:eval} shows, for each example, the total lines of code \marco{(not including whitespace)}, the number of lines used for annotations (i.e.,  specifications and proof \peter{annotations}), and the verification time in seconds (measured on an 8-core AMD Ryzen 6850U with 32GB of RAM running Ubuntu 22.04 on a warmed-up JVM, averaged over five runs), as well as the used data structure and abstraction. 
The examples range from \marco{29 to 140} lines of code, with a similar number of lines of specifications and proof annotations. \marco{Typical verification times are between 3 and 30 seconds.} Thus, we conclude that commutativity-based reasoning can be automated efficiently.
In the remainder of this section, we will discuss how different aspects of our technique \peter{allow us to verify the diverse set of} examples. 

\paragraph*{Precise action definitions}
For several examples, we can prove that the actions of all threads commute by carefully defining the legal actions.
For example, while puts on maps with the same keys do not commute in general, in the example Salary-Histogram, each put \emph{increments} the value for the key it is modifying (i.e., it increments the number of employees that fall within the specific salary range represented by the key). As a result, the updates commute.  
Similarly, the example Most-Valuable-Purchase iterates through purchase records and finds the highest price each user has ever paid: Here, each thread updates the map from user IDs to prices \emph{only if} its purchase value is greater than the one currently in the map, which again results in commuting map updates.

\paragraph*{Abstraction}
In half of our examples, we use an abstraction to relax the commutativity requirement. 
For example, 
we compute a list of employee names and salaries but leak only the mean salary, \marco{and} we collect a list of individual debts of a person but leak only the overall sum of their debt, not individual amounts or creditors.
In example Email-Metadata, we abstract a list to a multiset to prove that we may leak the list after sorting it, which eliminates the secret-dependent item order.
For the producer-consumer and pipeline examples, our abstraction does not return a view on the actual shared data, but on \emph{ghost data} added for verification purposes: It does not state that the current data in the queue is low, which is not of interest, since the final state of the queue will always be empty, but that the sequence of \emph{consumed} items is low (or, if there are multiple producers and consumers, the multiset view of that sequence).  
We also prove that the original version of \figref{initexample}, where each thread sets the shared variable to a different value, is correct if said value is not leaked.

\paragraph{Resource specifications}
The examples Sick-Employee-Names and Website-Visitor-IPs both \marco{add low values to sets} (names of sick employees and IP addresses of website visitors, respectively), but \marco{use} different set implementations. Since resource specifications abstract over concrete implementations of data structures, we can reuse the same resource spec for both examples.

\paragraph{Unique actions}
In three examples, we use unique actions to exploit that some actions are performed only by one thread, and thus their order is unaffected by thread interleavings.
In Sales-By-Region, different threads process data from different regions and add it to a shared map. Since the keys are region-specific, writes from different threads never conflict, and all updates commute.
For the single-producer-consumer and pipeline examples, we exploit that there is only one producer and one consumer thread per shared queue to prove that the sequence of consumed values is low; if there were multiple such threads (like in the multiple-producer-consumer example), the order of consumed items could be affected by secret data (and thus, only its multiset view is low).

\paragraph{Retroactive checking of action arguments}
For the multiple producer-consumer example, the number of items consumed by each individual consumer depends on scheduling and therefore potentially high data. However, the \emph{total} number of consumed items is low. Thus, checking that the number of performed consume actions is low \emph{after} joining all threads, when the overall number of consume actions is known, allows us to prove this example secure. 
In the pipeline example, a thread produces low data into a first queue, a middle thread consumes this data, transforms it, and produces the result into a second queue, from where a third thread consumes it. While all threads are executing, the middle thread does not know that the data it reads from the first queue is low; we learn that only once the first queue is unshared. 
Thus, we would not be able to prove that the middle thread fulfills the second queue's precondition (that produced data is low) while all threads are running, but we \emph{can} prove this precondition retroactively after unsharing the first queue.  

\paragraph{High branches}
Ca. half of our examples have secret-dependent timing due to branches on high data, and would thus be rejected by existing techniques, even if the attacker cannot observe timing.

\section{Related Work}\label{sec:rw}

Researchers have developed a plethora of type systems~\cite{smith}, static analyses~\cite{joana}, program transformations~\cite{cav21}, and program logics~\cite{covern,seccsl} to verify information flow security of concurrent programs, as well as multiple definitions of information flow security in this setting. 
Bisimulation-based properties~\cite{bisim} and observational determinism~\cite{obsdet} are properties of (sets of) traces, which assume that attackers can observe either low program variables or low events \emph{during} the execution of the program,
unlike our setting, where we assume that the attacker can observe \emph{only} the public output of the program.

In our setting, the standard property is non-interference (which is what our logic guarantees), or weaker versions of it like \emph{possibilistic}~\cite{poss} and \emph{probabilistic}~\cite{prob} non-interference,
which guarantee, respectively, that secret data does not influence either the set of possible final outputs or their probabilities. Since possibilistic non-interference is too weak in practice, most existing techniques target either traditional or probabilistic non-interference. 
As discussed previously, they achieve this goal by entirely preventing secret-dependent execution time differences, under the idealized assumption that only high branches can lead to such differences~\cite{smith,covern,seccsl,cav21,smith_prob_precise,veronica}. These techniques can be applied to realistic settings by also preventing other sources of timing differences, but preventing them entirely on standard hardware is complex and requires strong assumptions about compilers and hardware.
Our technique has the unique advantage of not requiring any reasoning about execution time in order to prove information flow security for output values, which makes it applicable independently of the used hardware. 
\marcon{In settings where it is possible to precisely reason about execution time, \logicName \emph{complements} existing techniques like SecCSL, and their relative completeness will vary from example to example (see \secref{limitations}).}
\marcon{In principle, it is also possible to verify concurrent programs by rewriting them as sequential programs  with non-deterministic scheduling and reasoning about them using techniques for sequential program. However, this approach would require explicitly considering the possible interleavings of all threads, which is not modular and does not scale in practice.}

Most of the aforementioned techniques either do not support a mutable heap (e.g. \cite{smith,covern} or work for programs using locks to protect shared memory like our logic (e.g.~\cite{seccsl,cav21}), but some recent logics target more complex settings, like fine-grained concurrency~\cite{seloc} or relaxed memory models~\cite{secrsl}. We believe that our approach also extends to fine-grained concurrency, since the general idea of making sure that concurrent changes commute also applies in this setting.

Some techniques prevent unwanted information leakage (including leakage as a result of internal timing channels) at \emph{runtime} by modifying the language runtime~\cite{DBLP:conf/post/VassenaSACRS19}, 
using information flow aware concurrency primitives~\cite{addressingcovert},
or by transforming the original program~\cite{DBLP:conf/asian/RussoHNS06}.
Such techniques can allow running programs that would be rejected by static techniques, but at the cost of requiring specific runtimes or affecting execution performance.

\marcon{
There are many use cases of commutativity in the literature, particularly in the context of parallelization. \logicName is the first technique that applies commutativity to information flow security in a concurrent setting.
}
\marcon{
\citet{DBLP:conf/ppdp/BentonKBH07} present a relational semantics that can \emph{prove} 
that two operations commute if the locations one reads and the other modifies do not overlap,
whereas we \emph{use} commutativity of actions to prove noninterference for programs that perform these actions. Additionally, we consider cases where different threads \emph{do} modify the same shared data, and thus \citet{DBLP:conf/ppdp/BentonKBH07}’s reasoning principles do not apply to our setting.}

\marcon{Some related work uses commutativity coupled with an abstraction, in particular, it uses data structures that expose a kind of abstract view and define commutativity on the basis of such a view. For example, \citet{DBLP:conf/pldi/KimR11} verify commutativity conditions in this context for specific data structures, and \citet{DBLP:conf/tacas/BansalKT18} automatically infer them. A lot of existing work \emph{uses} commutativity with such abstractions to reason about or transform programs. For example, \citet{DBLP:conf/ppopp/Golan-GuetaRSY15} use commutativity to parallelize sequential programs, \citet{DBLP:conf/pldi/DimitrovRVK14} use commutativity to detect races in traces of concurrent programs, and \citet{lucas} use it to prove serializability. \citet{DBLP:conf/fmcad/KraglQ21} use commutativity to simplify reasoning about concurrent programs, and \citet{pincus2022commutativity} uses it to parallelize code automatically and additionally infers commutativity conditions. All of them use a notion of abstraction that is more akin to our pure (mathematical) representations of the data structure (e.g., a partial function for a hash map). Our notion of “abstract view” is an additional layer of abstraction on top of the pure mathematical value which is application-specific and intentionally abstracts away high data (e.g., by further abstracting the partial function to its domain).
That is, existing work still uses and proves actual commutativity of data structure operations that ignores only internal implementation details of the data structure, whereas our abstractions allow operations that actually do not commute, if the result of this non-commutativity is not visible in public outputs.
Thus, given a data structure and its pure representation, we may use different abstractions for different usages in the same program.
}

\marcon{Finally, \citet{DBLP:journals/pacmpl/FarzanKP23} 
abstract \emph{statements} (e.g., a concrete assignment to a non-deterministic one) and reason about the commutativity of two abstracted statements \emph{on concrete states}, whereas we abstract states and check whether actions commute w.r.t. these abstracted states.
}

\section{Conclusion and Future Work}\label{sec:conclusion}

We have presented \logicName, a concurrent separation logic for verifying non-interference using abstract commutativity. We proved \logicName sound and showed that it can be automated efficiently and is able to verify common \peter{examples}.
For future work, we \marcon{plan to apply} reasoning based on (abstract) commutativity to other settings like fine-grained concurrency, and to explore when it is sound to  assume that shared data is low \emph{while} threads are still performing concurrent updates.

\bibliography{paper}

\ifarxiv

\appendix
\section{Appendix}

\subsection{Operational semantics}\label{app:opsem}
Fig.~\ref{fig:opsem} shows the operational small-step semantics of our language, which are standard and taken from \citet{cslsound}.

\begin{figure*}
\[
\footnotesize
\begin{array}{c}
\Inf[\textsc{Read}]{h(\mmeval{e}{s}) = v}
                   {\mmstepsto{\mmconf{\cread{x}{e}}{s}{h}}{\mmconf{\noop}{s[x \mapsto v]}{h}}} 
\quad
\Inf[\textsc{ReadA}]{\mmeval{e}{s} \notin \dom(h)}
                    {\mmstepsto{\mmconf{\cread{x}{e}}{s}{h}}{\abort}} \\
\ \\
\Inf[\textsc{Write}]{\mmeval{e_1}{s} \in \dom(h)}
                    {\mmstepsto{\mmconf{\cwrite{e_1}{e_2}}{s}{h}}{\mmconf{\noop}{s}{h[\mmeval{e_1}{s} \mapsto h\mmeval{e_2}{s}]}}} 
\quad
\Inf[\textsc{WriteA}]{\mmeval{e_1}{s} \notin \dom(h)}
                     {\mmstepsto{\mmconf{\cwrite{e_1}{e_2}}{s}{h}}{\abort}} \\
\ \\
\Inf[\textsc{Alloc}]{l \notin \dom(h)}
                    {\mmstepsto{\mmconf{\alloc{x}{e}}{s}{h}}{\mmconf{\noop}{s[x \mapsto l]}{h[l \mapsto \mmeval{e}{s}}}} \\ 
\ \\
\Inf[\textsc{Seq1}]{\mmstepsto{\mmconf{\seq{\noop}{c_2}}{s}{h}}{\mmconf{c_2}{s}{h}}} 
\quad
\Inf[\textsc{Seq2}]{\mmstepsto{\mmconf{c_1}{s}{h}}{\mmconf{c'_1}{s'}{h'}}}
                   {\mmstepsto{\mmconf{\seq{c_1}{c_2}}{s}{h}}{\mmconf{\seq{c'_1}{c_2}}{s'}{h'}}} \\ 
\ \\
\Inf[\textsc{SeqA}]{\mmstepsto{\mmconf{\seq{\abort}{c_2}}{s}{h}}{\abort}} 
\quad
\Inf[\textsc{Assign}]{\mmstepsto{\mmconf{\assign{x}{e}}{s}{h}}{\mmconf{\noop}{s[x \mapsto \mmeval{e}{s}]}{h}}} \\
\ \\
\Inf[\textsc{If1}]{\mmeval{b}{s}}
                   {\mmstepsto{\mmconf{\cond{b}{c_1}{c_2}}{s}{h}}{\mmconf{c_1}{s}{h}}} 
\quad
\Inf[\textsc{If2}]{\neg \mmeval{b}{s}}
                   {\mmstepsto{\mmconf{\cond{b}{c_1}{c_2}}{s}{h}}{\mmconf{c_2}{s}{h}}} \\
\ \\
\Inf[\textsc{Loop}]{\mmstepsto{\mmconf{\while{e}{c}}{s}{h}}{\mmconf{\cond{b}{\seq{c}{\while{e}{c}}}{\noop}}{s}{h}}} \\ 
\ \\
\Inf[\textsc{Par1}]{\mmstepsto{\mmconf{c_1}{s}{h}}{\mmconf{c'_1}{s'}{h'}}}
                   {\mmstepsto{\mmconf{\parc{c_1}{c_2}}{s}{h}}{\mmconf{\parc{c'_1}{c_2}}{s'}{h'}}}    
\quad
\Inf[\textsc{Par2}]{\mmstepsto{\mmconf{c_2}{s}{h}}{\mmconf{c'_2}{s'}{h'}}}
                   {\mmstepsto{\mmconf{\parc{c_1}{c_2}}{s}{h}}{\mmconf{\parc{c_1}{c'_2}}{s'}{h'}}} \\  
\ \\
\Inf[\textsc{Par3}]{\mmstepsto{\mmconf{\parc{\noop}{\noop}}{s}{h}}{\mmconf{\noop}{s}{h}}} \\    
\ \\
\Inf[\textsc{ParA1}]{\mmstepsto{\mmconf{c_1}{s}{h}}{\abort}}
                    {\mmstepsto{\mmconf{\parc{c_1}{c_2}}{s}{h}}{\abort}}   
\quad
\Inf[\textsc{ParA2}]{\mmstepsto{\mmconf{c_2}{s}{h}}{\abort}}
                   {\mmstepsto{\mmconf{\parc{c_1}{c_2}}{s}{h}}{\abort}} \\       
\ \\  
\Inf[\textsc{Atom}]{\mmstepstox{\mmconf{c}{s}{h}}{\mmconf{\noop}{s'}{h'}}}
                   {\mmstepsto{\mmconf{\atomic{c}}{s}{h}}{\mmconf{\noop}{s'}{h'}}} 
\quad
\Inf[\textsc{AtomA}]{\mmstepstox{\mmconf{c}{s}{h}}{\abort}}
                   {\mmstepsto{\mmconf{\atomic{c}}{s}{h}}{\abort}} \\ 
\ \\
\end{array}
\]
\vspace{-3mm}
\caption{Operational semantics.}\label{fig:opsem}
\end{figure*}

\section{Full Logic}\label{app:logic}

In this section, we provide additional definitions not shown in the main part of the paper.

\subsection{Additional Definitions}\label{app:logicdefs}
The sum of two guard heaps for a unique action $a_i$ is defined as follows: 

\begin{equation}
  \guheap_i \oplus \guheap'_i = 
  \begin{cases}
    \guheap_i &\text{ if } \guheap'_i = \bot \\
    \guheap'_i &\text{ if } \guheap_i = \bot \\
    \mathit{undefined}&\text{ otherwise} \\
  \end{cases}
\end{equation}

That is, at least one guard heap must have value $\bot$, and the sum simply contains the non-$\bot$ value (if any exists).

Addition of shared guard heaps computes the multiset union of the respective argument multisets of the two summands:

\begin{equation}
  \gsheap \oplus \gsheap' = 
  \begin{cases}
    \langle r + r', \mathit{args} \msadd \mathit{args}' \rangle&\text{ if }\gsheap = \langle r, \mathit{args} \rangle \wedge \gsheap = \langle r', \mathit{args}' \rangle \wedge r + r' \leq 1 \\
    \gsheap &\text{ if } \gsheap' = \bot \\
    \gsheap' &\text{ if } \gsheap = \bot \\
    \mathit{undefined}&\text{ otherwise} \\
  \end{cases}
\end{equation}

Finally, as stated before, permission heap addition is standard:

\begin{equation}
  (\pheap \oplus \pheap')(l) = 
  \begin{cases}
    \pheap(l) \oplus \pheap'(l)&\text{ if }l \in \funcdom{\pheap} \cap \funcdom{\pheap'} \\
    \pheap(l) &\text{ if }l \in \funcdom{\pheap} \wedge l \notin \funcdom{\pheap'} \\
    \pheap'(l) &\text{ if }l \in \funcdom{\pheap'}  \wedge l \notin \in \funcdom{\pheap}\\
  \end{cases}
\end{equation}
where
\begin{equation}
\langle r, v \rangle \oplus \langle r', v' \rangle =
\begin{cases}
  \langle r + r', v \rangle &\text{ if }r + r' \leq 1 \wedge v = v' \\
  \mathit{undefined}&\text{ otherwise}
\end{cases}
\end{equation}

\subsection{Additional Proof Rules}\label{app:sideconds}
\figref{fullrules} contains the proof rules not shown in the main part of the paper.

\begin{figure*}
\[
\footnotesize
\begin{array}{c}
\Inf[\textsc{Assign}]{\Gamma_{\bot} = \Gamma \Rightarrow x \notin \freevars{\Gamma}}
                     {\mmtriple{\Gamma_{\bot}}{P[e/x]}{\assign{x}{e}}{P}} 
\quad
\Inf[\textsc{New}]{x \notin \freevars{e}}
                  {\Gamma_{\bot} = \Gamma \Rightarrow x \notin \freevars{\Gamma}}
                  {\mmtriple{\Gamma_{\bot}}{\emp}{\alloc{x}{e}}{\pointsto{x}{1}{e}}}\\
\ \\
\Inf[\textsc{Read}]{x \notin \freevars{e_1, e_2}}
                   {\Gamma_{\bot} = \Gamma \Rightarrow x \notin \freevars{\Gamma}}
                   {\mmtriple{\Gamma_{\bot}}{\pointsto{e_1}{r}{e_2}}{\cread{x}{e}}{\pointsto{e_1}{r}{e_2} \ast x = e_2}}
\quad
\Inf[\textsc{Write}]{\mmtriple{\Gamma_{\bot}}{\pointsto{e_1}{1}{\_}}{\cwrite{e_1}{e_2}}{\pointsto{e_1}{1}{e_2}}}\\
\ \\
\Inf[\textsc{If1}]{\mmtriple{\Gamma_{\bot}}{P \wedge b}{c_1}{Q}}
								 {\mmtriple{\Gamma_{\bot}}{P \wedge \neg b}{c_2}{Q}}
                 {\mmtriple{\Gamma_{\bot}}{P \wedge \below{b}{\llow}}{\cond{b}{c_1}{c_2}}{Q}}\\
\ \\
\Inf[\textsc{If2}]{\mmtriple{\Gamma_{\bot}}{P \wedge b}{c_1}{Q}}
								 {\mmtriple{\Gamma_{\bot}}{P \wedge \neg b}{c_2}{Q}}
								 {\unaryass{Q}}
                 {\mmtriple{\Gamma_{\bot}}{P}{\cond{b}{c_1}{c_2}}{Q}}\\
\ \\
\Inf[\textsc{While1}]{\mmtriple{\Gamma_{\bot}}{P \wedge b}{c_l}{P \wedge \below{b}{\llow}}}
                 {\mmtriple{\Gamma_{\bot}}{P \wedge \below{b}{\llow}}{\while{b}{c_l}}{P \wedge \neg b}}
\quad
\Inf[\textsc{While2}]{\mmtriple{\Gamma_{\bot}}{P \wedge b}{c_l}{P}}
								 {\unaryass{P}}
                 {\mmtriple{\Gamma_{\bot}}{P}{\while{b}{c_l}}{P \wedge \neg b}}\\
\ \\
\Inf[\textsc{Seq}]{\mmtriple{\Gamma_{\bot}}{P}{c_1}{R}}
									{\mmtriple{\Gamma_{\bot}}{R}{c_2}{Q}}
                  {\mmtriple{\Gamma_{\bot}}{P}{\seq{c_1}{c_2}}{Q}}
\quad
\Inf[\textsc{Skip}]{\mmtriple{\Gamma_{\bot}}{P}{\noop}{P}} \\
\ \\
\Inf[\textsc{Par}]{\mmtriple{\Gamma_{\bot}}{P_1}{c_1}{Q_1}}
									{\mmtriple{\Gamma_{\bot}}{P_2}{c_2}{Q_2}}
									{\freevars{P_1, c_1, Q_1} \cap \modset{c_2} = \emptyset}
									{\freevars{P_2, c_2, Q_2} \cap \modset{c_1} = \emptyset}.
									{\Gamma_{\bot} = \Gamma \Rightarrow \freevars{\Gamma} \cap \modset{c_1, c_2} = \emptyset}
									{P_1\text{ is precise or }P_2\text{ is precise}}
                  {\mmtriple{\Gamma_{\bot}}{P_1 \ast P_2}{\parc{c_1}{c_2}}{Q_1 \ast Q_2}}\\
\ \\
\Inf[\textsc{Cons}]{P \Rightarrow P'}
									 {\mmtriple{\Gamma_{\bot}}{P'}{c}{Q'}}
									 {Q' \Rightarrow Q}
                   {\mmtriple{\Gamma_{\bot}}{P}{c}{Q}}
\quad
\Inf[\textsc{Frame}]{\freevars{R} \cap \modset{c} = \emptyset}
									 {\mmtriple{\Gamma_{\bot}}{P}{c}{Q}}.
									 {P\text{ is precise or }R\text{ is precise}}
                   {\mmtriple{\Gamma_{\bot}}{P \ast R}{c}{Q \ast R}}\\
\ \\
\Inf[\textsc{Exists}]{x \notin \freevars{c}}{\mathit{unambiguous}(P, x)}
                   {\Gamma_{\bot} = \Gamma \Rightarrow x \notin \freevars{\Gamma}}
									 {\mmtriple{\Gamma_{\bot}}{P}{c}{Q}}
                   {\mmtriple{\Gamma_{\bot}}{\exists x \ldotp P}{c}{\exists x \ldotp Q}} \\
\end{array}
\]
\vspace{-3mm}
\caption{Remaining proof rules. The rules for conditionals and loops have been adjusted for information flow reasoning. All other rules are as in standard CSL. As before, we use $\freevars{P, e}$ to denote the free variables in assertions and expressions (\freevars{\Gamma} is the set of free variables in the invariant $I$), and $\modset{c}$ for the set of variables modified by command $c$.}\label{fig:fullrules}
\end{figure*}

As stated before, for basic statements, they are essentially identical to standard CSL (except for some precision \marcon{and non-ambiguity} requirements, see below).

The remaining main difference between our logic and standard CSL is in the rules for conditionals and loops. 
There are two rules for conditionals: The first can be used if the condition can be shown to be low, i.e., both executions will use the same branch. If this is not the case, the second rule must be used, which limits the postcondition $Q$ to be unary. This restriction prevents indirect flows; for example, for the command $\cond{h}{\assign{x}{1}}{\assign{x}{0}}$, where $h$ is high, this rule does not allow proving the postcondition $\below{x}{\llow}$, which holds for each branch individually, but does not hold if one execution executes the then-branch and the other execution chooses the else-branch.

The same approach is used for loops: If the loop condition can be shown to be low initially and to remain low after each loop iteration, one can use a relational (non-unary) loop invariant. Alternatively, if the loop condition may be high, we restrict the invariant to be unary\footnote{It is possible to generalize these rules to a more complete rule that allows using a relational invariant for some \emph{prefix} of the loop iterations~\cite{cartesian}, but we do not do this here because loop handling is orthogonal to our main contribution.}.

\subsection{Precision \marcon{and Non-Ambiguity} Side Conditions}

\marcon{Rule \textsc{Exists} has a side condition $\mathit{unambiguous}(P, x)$, which intuitively states that $P$ uniquely constrains the value of $x$ in any given pair of states~\cite{DBLP:journals/pacmpl/DardinierMS22}:}

\begin{definition}
$\mathit{unambiguous}(P, x)$ holds for some assertion $P$ and some variable $x$ iff, for all $s_1, \gheap_1, s_2, \gheap_2$ and all values $v_1, v_2, v'_1, v'_2$, if $(s_1[x \mapsto v_1], \gheap_1), (s_2[x \mapsto v_2], \gheap_2) \models P)$ and $(s_1[x \mapsto v'_1], \gheap_1), (s_2[x \mapsto v'_2], \gheap_2) \models P)$, then $v_1 = v'_1$ and $v_2 = v'_2$.
\end{definition}

Similarly, rules \textsc{Frame}, \textsc{Par}, \textsc{Share}, \textsc{AtomicShr}, and \textsc{AtomicUnq} contain precision requirements that do not exist in their standard CSL equivalents. While we believe these restrictions can be lifted to something weaker, \emph{some} restrictions on the assertions are necessary in our setting; existing logics do not require them because they do not reason about pairs of executions with potentially different schedules.

The actual condition we require at different points in the proof (e.g., in the frame rule for $P_1 = P$ and $P_2 = R$) is the following:
If
$\gheap, \gheap’ \models P_1 \ast P_2$ and 
$\gheap = \gheap_1 \oplus \gheap_2$, where $\gheap_1, \gheap_1 \models P_1$ and $\gheap_2, \gheap_2 \models P_2$, and similarly
$\gheap' = \gheap'_1 \oplus \gheap'_2$, where $\gheap'_1, \gheap'_1 \models P_1$ and $\gheap'_2, \gheap'_2 \models P_2$
then we must have
$\gheap_1, \gheap'_1 \models P_1$ and $\gheap_2, \gheap'_2 \models P_2$.

While this condition does not hold in general, it holds if $P_1$ or $P_2$ is precise, which is why we introduce the precision requirements.
These requirements could be omitted throughout the logic by instead restricting the assertion language (in particular, allowing only specific forms of disjunctions and existential quantification that always fulfill the condition); this is what we do in our implementation, meaning that no such side condition has to be checked there. In our experience, restricting the assertion language does not limit the expressiveness of assertions in a practically meaningful way.

\subsection{Noguard Side Conditions}
The noguard restrictions in the atomic rules are required as a result of our definition of safety (Definition~\ref{def:safety}), which is essentially trivially true for states with an empty resource context but non-empty guard states, since generally guards are meaningful only if there is a shared resource. However, in the atomic-rules, if $P$ were to contain a guard assertion, then its premise would be a Hoare triple with an empty resource context and a guard in the precondition. Thus, an induction proof would fail at this point, since the premise does not guarantee anything about the execution of $c$. 

To prevent this problem without having to complicate the definition of safety only for this one case, we enforce that $P$ and $Q$ contain no guards here. In practice, this side condition is not restrictive; for any atomic statement, all guards held by the current thread that are not used for the action performed by the atomic statement can be framed around it, and it is never necessary to have another guard in $P$ to verify the body of the atomic statement.

\section{Details on Soundness}\label{app:soundness}
Here, we provide our formal definition of the safety predicate $\mathit{safe}_n$. 
We use $\gheapfg$ to range over extended heaps whose guard states are $\bot$ and whose permission heaps have full permission of every location in their domain, and $\gheapf$ to range over extended heaps whose permission heaps have full permission of every location in their domain (but who may have arbitrary guard states):

\begin{definition}\label{def:safety}
$\mathit{safe}_0(\Gamma_{\bot}, c, \langle s, \gheap \rangle, \Sigma)$ holds always.

$\mathit{safe}_{n+1}(\bot, c, \langle s, \gheap \rangle, \Sigma)$ holds iff
\begin{enumerate}
  \item If $c = \noop$ then $\langle s, \gheap \rangle \in \Sigma$.
  \item $c$ will never abort from any state $\langle s, \normalize{\gheapfg} \rangle$ s.t.\ $\gheapfg = \gheap \oplus \gheap_f$ for some $\gheap_f$. 
  \item For all $\gheapfg, \gheap_f, c', s', \nheap'$, where $\gheapfg = \gheap \oplus \gheap_f$, if $c, \langle s, \normalize{\gheapfg} \rangle \rightarrow c', \langle s', \nheap' \rangle$, then there is some $\gheap'$ and some $\gheapfg'$ s.t.\ $\gheap' = \normalize{\gheapfg'}$ and $\gheapfg' = \gheap' \oplus \gheap_f$, and $\mathit{safe}_{n}(\bot, c', \langle s', \gheap' \rangle, \Sigma)$.
\end{enumerate}

$\mathit{safe}_{n+1}(\Gamma, c, \langle s, \gheap \rangle, \Sigma)$ holds iff
\begin{enumerate}
  \item If $c = \noop$ then $\langle s, \gheap \rangle \in \Sigma$.
  \item $c$ will never abort from any state $\langle s, \normalize{\gheapf} \rangle$ s.t.\ $\gheapf = \gheap \oplus \gheap_I \oplus \gheap_f$ for some $\gheap_f, \gheap_I$ s.t.\ $\gheap_I$ satisfies $I(v)$ for some $v$ and $\gheapf$ is consistent with $\Gamma$ from some $v_0$. 
  \item For all $\gheapf, \gheap_f, \gheap_I, c', s', \nheap', v_0$, where $\gheapf = \gheap \oplus \gheap_I \oplus \gheap_f$ and $\gheap_I$ satisfies $I(v)$ for some $v$ and $\gheapf$ is consistent with $\Gamma$ from $v_0$, if $c, \langle s, \normalize{\gheapf} \rangle \rightarrow c', \langle s', \nheap' \rangle$, then there is some $\gheap', \gheap'_I, \gheapf'$ s.t.\ $\gheap'_I$ fulfills $I(v')$ for some $v'$, $\gheapf$ is consistent with $\Gamma$ from $v_0$, $\gheap' = \normalize{\gheapf'}$ and $\gheapf' = \gheap' \oplus \gheap'_I \oplus \gheap_f$, and $\mathit{safe}_{n}(\Gamma, c', \langle s', \gheap' \rangle, \Sigma)$.
\end{enumerate}
\end{definition}

That is, the two non-trivial cases where $n \neq 0$ both contain three almost identical clauses:
The first requires that if the state is final, it is contained in $\Sigma$; 
the second requires that the program will not abort from the current state extended with an arbitrary frame $\gheap_f$ and, if there is a shared resource, a resource heap $\gheap_I$ that satisfies the invariant. The extension by an arbitrary heap is standard in CSL and necessary to prove the frame rule; additionally, here, it is used to ``complete''  the permission heap to contain full permissions for every location in its domain. Additionally, recall that, since the operational semantics is defined on ordinary heaps without permissions or guard states, we have to normalize the extended heap first before talking about executions from this state.
The third clause requires that performing a single step from the given state (again, extended with an arbitrary frame $\gheap_f$ and, if there is a shared resource, a resource heap $\gheap_I$, and normalized to get a normal heap) preserves the frame, and that recursively, the next $n$ steps will also be safe from the new state. Additionally, in the second case (where there is a shared resource), it requires that the state remains consistent w.r.t.\ some initial value $v_0$, i.e., that the state of the resource heap continues to be a possible result of applying the actions with the arguments specified in the guards to the initial value $v_0$.

\section{Partial Actions}\label{app:partial}

Our actions have to be total w.r.t.\ the value of the shared resource. In this section, we will explain the reason for this requirement, and how restrictive it is in practice. 
Consider an example consisting of two \emph{producer} threads which produce some data and put it in a shared queue, and a single \emph{consumer} thread that reads the data from the queue and subsequently processes it in some way. For illustration purposes, we limit the size of the buffer to one. 

Both producing and consuming an item are now possible only in certain states of the shared resource: The consumer can  consume an item only when the queue is non-empty, and the producers can  produce an item only when it is empty (since we limited the size of the buffer to 1). Thus, in such an implementation, threads would have to block before acquiring the shared resource until they can execute their action; this could be supported e.g. using a command $\code{atomic }c\code{ when }e$ that waits until $e$ is true before executing $c$ atomically. As a result, the natural definitions of the produce and consume actions are \emph{partial} functions, which is not allowed by our framework.

\begin{figure}
\includegraphics[width=0.8\textwidth]{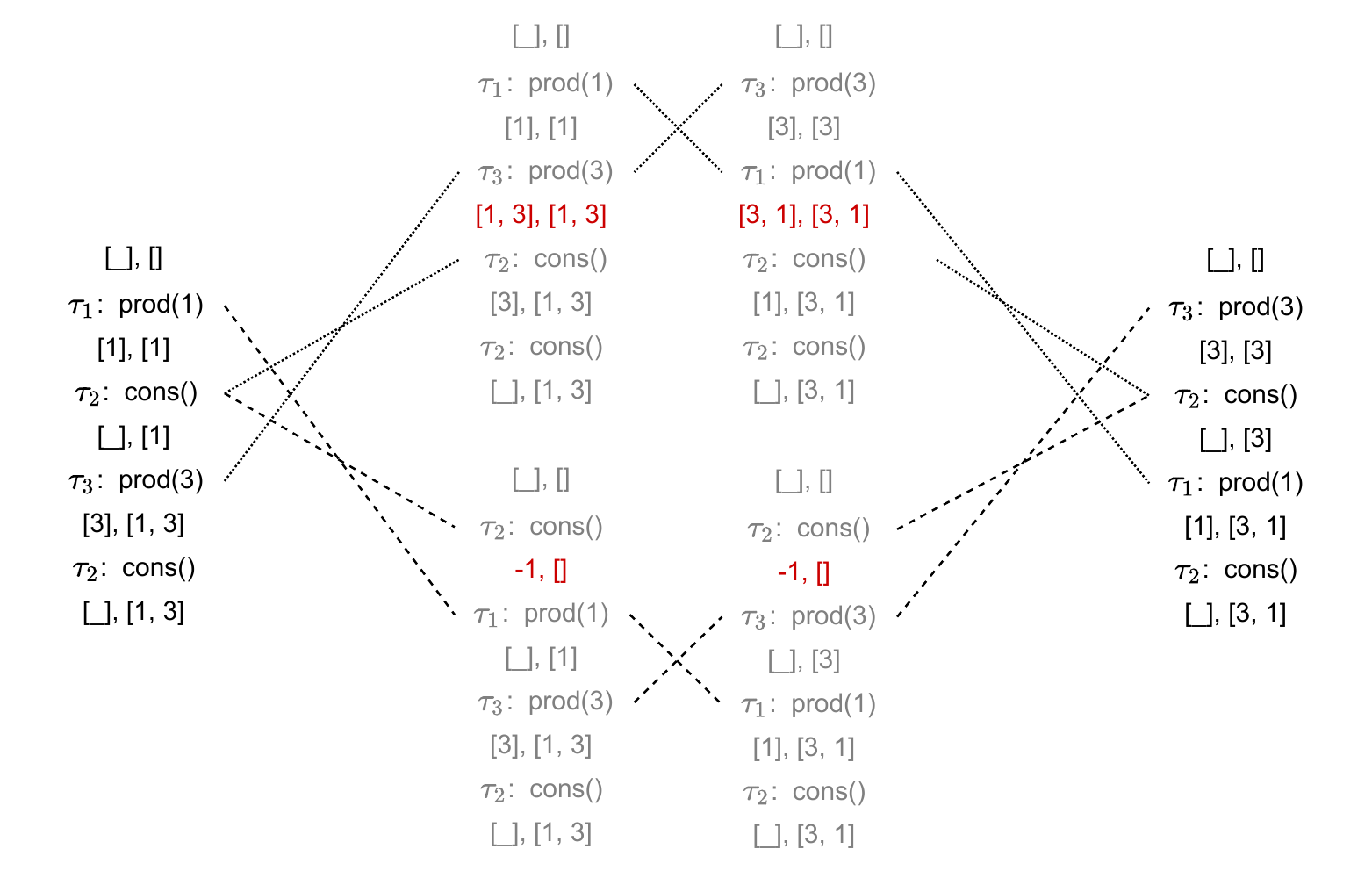}
\vspace{-3mm}
\caption{Two executions of a producer-consumer example (left and right). When using pairwise swaps to permute one into the other, some resulting intermediate states (marked in red) are impossible (single-element buffers with more than one or less than zero elements).}\label{fig:buffer}
\end{figure}

The reason for this is that, if one were to allow partial actions and adapt the abstract commutativity requirement s.t.\ it  requires actions to commute only when all involved function applications are defined, our technique would become unsound.
as illustrated by the example in \figref{buffer}, which shows two possible traces of this program (left and right, ignore the rest of the figure for now), where both producer threads produce only a single item (thread $\tau_1$ produces the number 1, and thread $\tau_3$ produces the number 3, and the consumer thread $\tau_2$ consumes both), starting from an empty buffer. 
In the execution on the left, $\tau_1$ produces its item first, whereas in the execution on the right, $\tau_3$ is faster. The depicted states have the form $l, s$, where $l$ is the buffer (we write $[\_]$ when the buffer is empty and $[v]$ for a buffer containing value $v$), and $s$ is the sequence of produced values. Note that the multiset of the produced items is identical in both executions, whereas the sequence of produced items is not.

In our scenario, the two executions, left and right, are the \emph{only} possible executions of our program, all other schedules are impossible: Since the buffer is initially empty, the first action \emph{has to} be a produce-action (the consumer would block until the queue is non-empty), and after a produce-action has happened, the buffer is full, meaning that the next action can only be a consume-action. However, as a result, if produce and consume are partial functions, it is impossible to use pairwise swaps to permute one execution to the other using pairwise swaps: In fact, no swaps are possible at all. 

However, the soundness of our technique depends on the ability to permute one sequence of actions into the other via pairwise swaps, since our commutativity criterion only requires pairwise commutativity for all actions.
The example illustrates this:
A weakened abstract commutativity criterion where  any two actions $f_1$ and $f_2$ have to commute only when both orders of applications $f_1(f_2(x))$ and $f_2(f_2(x))$ are defined is unsound here, since it is trivially fulfilled for \emph{any} abstraction, since it is never the case that both orders are defined: In particular, $\mathit{cons}(\mathit{cons}(x))$ and $\mathit{prod}(\mathit{prod}(x))$ are never defined, and for any $x$, only one of $\mathit{prod}(\mathit{cons}(x))$ and $\mathit{cons}(\mathit{prod}(x))$ is defined. As a result, no check would prevent us from, for example, declaring the entire sequence of produced items (as opposed to its multiset abstraction) to be low, which would be unsound.

Thus, we require actions to be total functions on the resource value, but users can still support examples like the shown one by \emph{artificially} making actions total, possibly by modifying the type of the shared state in such a way that it can represent orders of actions that cannot happen in reality. In \figref{buffer} (and in the producer-consumer example in our evaluation), we did this by 1) allowing the queue to be longer than one element, and 2) allowing it to contain less than zero elements (in which case we only track the (negative) length of the sequence). That is, we define consuming in such a way that consuming from an empty queue leaves its state as $-1$ (and consuming from a queue whose size is already $-n$ results in a state $-(n+1)$). Conversely, producing when the queue is full adds an item to the queue, producing an item when a queue has state $-1$ results in an empty queue, and producing when the state is $-(n +1)$ results in state $-n$. We show the resulting resource specification, with its ghost state, in \figref{prodconsspec1}.

\begin{figure}
	\begin{align*}
\locktype{\textsc{PC}}{v} &\equiv \langle Either[Nat, Seq[Int]], Seq[Int] \rangle \\
\lockabstr{\textsc{PC}}{\langle v_1, v_2 \rangle} &\equiv \seqtoms{v_2} \\
	\sharedactions_{\textsc{PC}} &\equiv \{ \textsc{Prod} \} \\
	\uniqueactions_{\textsc{PC}} &\equiv \{ \textsc{Cons} \}\\
 	f_{\textsc{Prod}}(v, a) &\equiv \begin{cases} 
 	\langle \rightv{xs ++ [a]}, s ++ [a] \rangle & \text{ if } v = \langle \rightv{xs}, s \rangle \\
 	\langle \rightv{[]}, s ++ [a]  \rangle & \text{ if } v = \langle \leftv{-1}, s \rangle  \\
 	\langle \leftv{-n} , s ++ [a] \rangle & \text{ if } v = \langle \leftv{-(n+1)}, s \rangle
 	\end{cases} \\
 	\lockpre{\textsc{Prod}}{a} &\equiv \below{a}{\llow} \\
 	f_{\textsc{Cons}}(v, a) &\equiv \begin{cases} 
 	\langle \rightv{xs}, s \rangle & \text{ if } v = \langle \rightv{x :: xs}, s \rangle \\
 	\langle \leftv{-1}, s \rangle & \text{ if } v = \langle \rightv{[]}, s \rangle  \\
 	\langle \leftv{-(n + 1)} , s \rangle & \text{ if } v = \langle \leftv{-n}, s \rangle
 	\end{cases} \\
 	\lockpre{\textsc{Cons}}{a} &\equiv \trueval \\
	\end{align*}
	\vspace{-3mm}
	\caption{Producer-consumer resource specification. The second part of the tuple represents the sequence of produced items, whereas the first part represents the current contents of the buffer/queue. We mark real queue contents as $\mathit{right}$ and queues with negative contents as $\mathit{left}$. The abstraction function returns the multiset view of the sequence of produced items.}\label{fig:prodconsspec1}
\end{figure}

We then prove that these definitions of the produce- and consume-actions commute modulo our abstraction, which they do when using the multiset-abstraction (but not when using the set of produced items directly). Now, conceptually, it is possible to permute the execution on the left into the one on the right using swaps of neighboring actions, as shown in \figref{buffer} in the middle (we show two possible sequences of swaps that have the same result). Both of these swaps lead to intermediate states that cannot occur in any real run (highlighted in red), but the existence of these sequences, along with our commutativity criterion, guarantees that each such swap will leave (the abstract view of) the final state unchanged; therefore, we can conclude that both real executions, left and right, will lead to end results that are equal modulo abstraction.

Crucially, \emph{any} way of making the action functions total is valid, as long as the functions result in a resource specification that fulfills our validity criteria. In practice, we have found that there are a few typical patterns that allow one to accommodate most relevant cases: For example, to model reading from an empty data structure, it is generally sufficient to store its (negative) length instead of its contents, as in our example.  

\section{\toolName example} \label{app:hyperviper}
Here, we show a slightly simplified version of the relevant parts of an example similar to \figref{mapkeyexample} encoded into \toolName.
First, users have to declare resource specifications that define allowed actions as well as the abstract view of the resource:
\begin{lstlisting}
// Resource specification
specType MapLock {
  type Map[Int, Int]
  invariant(l, v) = [l.lockMap |-> ?mp && isMap(mp) && v == mapValue(mp)]
  alpha(v): Set[Int] = keys(v)

  actions = [(Put, Pair[Int, Int], duplicable)]
  
  action Put(v, arg)
    requires low(fst(arg))
  { (put(v, fst(arg), snd(arg))) }

  noLabels = N()
}
\end{lstlisting}

Then, the program itself is verified using said resource specifications.
We first show the main method which spawns a variable number of worker threads (using fork and join statements).
Guard assertions \code{sguard} correspond directly to the ones shown in this paper, but \toolName uses a function \code{sguardArgs} to refer to the argument of the current guard.
Note that guards are explicitly associated with resources, since  \toolName supports multiple shared resource in one program, and that shared action guards have to be explicitly split (in the first loop) and merged (in the second loop) using ghost statements.
Sometimes, it is necessary to assert lemmas about e.g. multisets in order to prove that specific splits are allowed, since the verifier cannot prove these completely without user guidance.

\begin{lstlisting}
// general definitions
function N(): Int
    ensures result > 1

function nTargets(): Int

// class TargetInfo
field addr: Int
field reason: Int

define targetInfo(r) (
    [r.addr |-> _] && [r.reason |-> ?r && low(r)]
)

field lockMap : Ref

method main(inputs: Seq[Seq[Ref]])
  requires lowEvent && low(|inputs|) && |inputs| == N()
  requires forall ip: Int :: ip >= 0 && ip < N() ==> low(|inputs[ip]|)
  requires forall ip: Int, jp: Int :: ip >= 0 && ip < N() && jp >= 0 && jp < |inputs[ip]| 
           ==> targetInfo(inputs[ip][jp])
{
  var m : Ref
  m := createMap()
  var l : Lock
  l := new(lockMap)
  l.lockMap := m
  share[MapLock](l, empty())
  var i : Int := 0
  var threads : Array
  threads := emptyArray(|inputs|)
  while (i < |inputs|)
    invariant i >= 0 && i <= |inputs| && low(i)
    invariant forall j : Int :: j >= 0 && j < |inputs| && j >= i 
              ==> [loc(threads, j).arr_loc |-> _]
    invariant forall j : Int :: j >= 0 && j < i 
              ==> joinable(loc(threads, j), inputs[j], l, j)
    invariant sguard[MapLock,Put](l, intervalSet(i, N())) && 
              sguardArgs[MapLock, Put](l, intervalSet(i, N())) == Multiset[Pair[Int, Int]]()
    invariant forall ip: Int, jp: Int :: ip >= i && ip < N() && jp >= 0 && jp < |inputs[ip]| 
              ==> targetInfo(inputs[ip][jp])
  {
    var t: Thread
    assert Set(i) union intervalSet(i+1, N()) == intervalSet(i, N())
    split[MapLock,Put](l, Set(i), intervalSet(i+1, N()), 
                       Multiset[Pair[Int, Int]](), Multiset[Pair[Int, Int]]())
    t := fork worker(inputs[i], l, i)
    loc(threads, i).arr_loc := t
    fold joinableArrayEntry(loc(threads, i), inputs[i], l, i)
    i := i + 1
  }
  
  i := 0
  while (i < |inputs|)
    invariant i >= 0 && i <= |inputs| && low(i)
    invariant forall j : Int :: j >= 0 && j < |inputs| && j >= i 
              ==> joinable(loc(threads, j), inputs[j], l, j)
    invariant forall j : Int :: j >= 0 && j < i ==> [loc(threads, j).arr_loc |-> _] 
    invariant sguard[MapLock,Put](l, intervalSet(0, i)) && 
              allPre[MapLock, Put](sguardArgs[MapLock,Put](l, intervalSet(0, i)))
  {
    unfold joinableArrayEntry(loc(threads, i), inputs[i], l, i)
    join[worker](loc(threads, i).arr_loc)
    assert Set(i) union intervalSet(0, i) == intervalSet(0, i + 1)
    merge[MapLock, Put](l, Set(i), intervalSet(0, i))
    i := i + 1
  }
  unshare[MapLock](l)

  var res: Seq[Int]
  res := toSeq(keys(mapValue(l.lockMap)))
  print(res)

}
\end{lstlisting}

Finally, for the worker method, we prove that the invariant that all arguments of the \code{Put} action fulfill its precondition is maintained. 
The \code{with}-command, which acquires a shared resource and then atomically executes its body, has to be annotated with the action it is supposed to perform.
\begin{lstlisting}
method worker(inputs: Seq[Ref], l: Lock, lbl: Int)
  requires lowEvent && low(|inputs|) && sguard[MapLock,Put](l, Set(lbl)) && 
           sguardArgs[MapLock,Put](l, Set(lbl)) == Multiset[Pair[Int, Int]]()
  requires forall jp: Int :: jp >= 0 && jp < |inputs| ==> targetInfo(inputs[jp])
  ensures sguard[MapLock,Put](l, Set(lbl)) && 
          allPre[MapLock, Put](sguardArgs[MapLock,Put](l, Set(lbl)))
{
  var i : Int := 0
  while (i < |inputs|) 
    invariant i >= 0 && i <= |inputs| && low(i) && sguard[MapLock,Put](l, Set(lbl)) && 
              allPre[MapLock, Put](sguardArgs[MapLock,Put](l, Set(lbl)))
    invariant forall jp: Int :: jp >= 0 && jp < |inputs| ==> targetInfo(inputs[jp])
  {
    var k: Int
    var v: Int
    k := inputs[i].addr
    v := inputs[i].reason
    with[MapLock] l performing Put(pair(k, v)) at lbl {
        var tmp : Map[Int, Int]
        tmp := mapValue(l.lockMap)
        mapPut(l.lockMap, k, v)
    }
    i := i + 1
  }
}
\end{lstlisting}
\else \fi

\end{document}